\documentclass[
    aps,
    prl,
    reprint,
    superscriptaddress,
    longbibliography
]{revtex4-2}
\usepackage{appendix}
\usepackage{dsfont}
\usepackage{bm}
\usepackage{enumerate}
\usepackage{graphicx}
\usepackage{dcolumn}
\usepackage{bm}

\usepackage{amssymb}
\usepackage{amsmath}
\usepackage{booktabs}
\usepackage{braket}
\usepackage[caption = false]{subfig}
\usepackage{bbm}
\usepackage{tikz}
\usetikzlibrary{patterns}
\usetikzlibrary{decorations.pathreplacing}
\tikzset{
    cross/.pic = {
    \draw[rotate = 45] (-#1,0) -- (#1,0);
    \draw[rotate = 45] (0,-#1) -- (0, #1);
    }
}
\usetikzlibrary {patterns.meta}
\pgfdeclarepattern{
      name=hatch,
      parameters={\hatchsize,\hatchangle,\hatchlinewidth},
      bottom left={\pgfpoint{-.1pt}{-.1pt}},
      top right={\pgfpoint{\hatchsize+.1pt}{\hatchsize+.1pt}},
      tile size={\pgfpoint{\hatchsize}{\hatchsize}},
      tile transformation={\pgftransformrotate{\hatchangle}},
      code={
        \pgfsetlinewidth{\hatchlinewidth}
        \pgfpathmoveto{\pgfpoint{-.1pt}{-.1pt}}
        \pgfpathlineto{\pgfpoint{\hatchsize+.1pt}{\hatchsize+.1pt}}
        \pgfpathmoveto{\pgfpoint{-.1pt}{\hatchsize+.1pt}}
        \pgfpathlineto{\pgfpoint{\hatchsize+.1pt}{-.1pt}}
        \pgfusepath{stroke}
      }
    }

\tikzset{
      hatch size/.store in=\hatchsize,
      hatch angle/.store in=\hatchangle,
      hatch line width/.store in=\hatchlinewidth,
      hatch size=3pt,
      hatch angle=0pt,
      hatch line width=.5pt,
    }

\newcommand{\ee}{{\rm e}}
\newcommand{\ii}{{\rm i}}

\tikzset{every picture/.style={line width=0.75pt}} 

\usepackage[normalem]{ulem} 

\begin{document}

\preprint{APS/123-QED}

\title{Unravelling the Li--Haldane Conjecture with the Projected Ensemble}

\author{Daniel Spasic-Mlacak}
\affiliation{T.C.M Group, Cavendish Laboratory, University of Cambridge, J.J. Thompson Avenue, Cambridge CB3 0US, United Kingdom}

\author{Qi Camm Huang}
\affiliation{Department of Physics, National University of Singapore, Singapore 117551}
\affiliation{Centre for Quantum Technologies, National University of Singapore, Singapore 117543}

\author{Wen Wei Ho}
\affiliation{Department of Physics, National University of Singapore, Singapore 117551}
\affiliation{Centre for Quantum Technologies, National University of Singapore, Singapore 117543}

\author{Nigel R. Cooper}
\affiliation{T.C.M Group, Cavendish Laboratory, University of Cambridge, J.J. Thompson Avenue, Cambridge CB3 0US, United Kingdom}

\date{\today}

\begin{abstract}
The entanglement spectra of fractional quantum Hall states contain universal fingerprints of their underlying topological order, as posited by the Li--Haldane conjecture. In this work, we uncover a finer universal structure within the entanglement spectra unravelled by projective measurements.
Concretely, we study the \textit{projected ensemble}---the collection of quantum states on a subsystem conditioned on measurement outcomes of its complement---of fractional quantum Hall states.
We find that this ensemble exhibits a hidden hierarchy inside 
the Li--Haldane edge manifold: by conditioning on measurement outcomes, the entanglement spectrum's support is split into measurement-dependent sectors whose ranks we demonstrate are fixed by conformal field theory counting, an observation we dub the {\it measurement-resolved} Li--Haldane conjecture.
For the non-Abelian Moore--Read state, this hierarchy is particularly rich: each parity-resolved edge manifold contains internal subspaces whose dimensions reproduce the conformal field theory counting of the opposite-parity sector. This structure persists even in realistic Coulomb-interacting ground states, establishing the projected ensemble as a sharp new probe of topological order beyond what the entanglement spectrum alone can detect.
\end{abstract}

\maketitle
{\it Introduction.}---Topological phases of matter cannot be characterized by local order parameters and instead require intrinsically non-local probes of quantum correlations. Quantum entanglement has emerged as one of the most powerful tools for diagnosing topological order, revealing universal properties inaccessible to conventional observables \cite{kitaev2006topological,levin2006detecting,laflorencie2016quantum}. Among these developments, the entanglement spectrum introduced by Li and Haldane demonstrated 
that the spectrum of the reduced density matrix (RDM) contains detailed information about the edge conformal field theory (CFT) describing the underlying topological phase \cite{Li-08-EntanglementSpectrum}. This observation established a deep connection between bulk entanglement and edge physics and has since become one of the central organizing principles in the study of fractional quantum Hall (FQH) states and other topological phases \cite{chandran2011bulk,sterdyniak2011extracting,qi2012general,dubail2011real}.

For FQH wavefunctions, the Li--Haldane conjecture states that the low-lying entanglement spectrum exhibits the same level counting as the corresponding chiral edge CFT \cite{Li-08-EntanglementSpectrum}. For model states, the rank of each block of the RDM is bounded by the CFT counting in the corresponding particle-number and angular-momentum sector, with the bound becoming saturated in the thermodynamic limit \cite{chandran2011bulk}. While this establishes that the RDM contains useful physical information about the edge theory of an FQH state through the dimensionality of the entanglement subspace it defines, it is an open question whether the structure of quantum states within the entanglement subspace itself can reveal additional universal information.

\begin{figure}[t]
\centering
\includegraphics[width=3.42in]{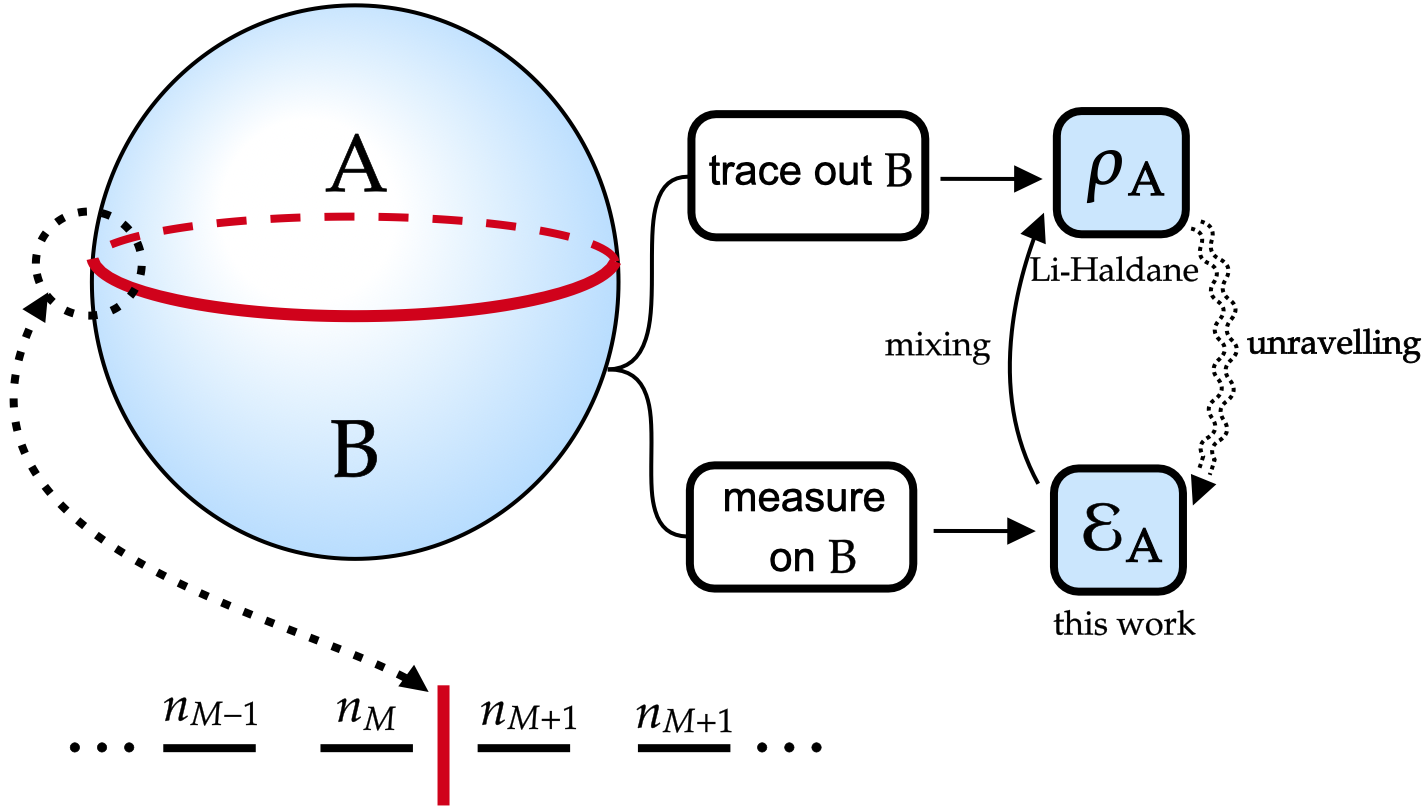}\
\caption{Schematic diagram of our setup: the FQH system on a sphere is bipartitioned into subsystems $\mathrm{A}$ and $\mathrm{B}$ using an orbital cut, with the angular momentum $\Delta L_{\mathrm{A}}$ and particle number $N_{\mathrm{A}}$ being fixed. By taking measurements on part of the subsystem, we construct the projected ensemble $\mathcal{E}_{\mathrm{A}}$ in the occupation number basis. }
\label{schematicPEonsphere}
\end{figure}

Motivated by recent studies of deep thermalisation~\cite{cotler_emergent_2023, ho2022exact, mark_maximum_2024, lucas_fermions_2023, liu_cv_2024, bejan_matchgate_2025, chang_deep_2025, liu_coherence-induced_2025, mcginley2025scrooge, mok_nature_2026} in quantum many-body dynamics---where finer structure of quantum states on a local subsystem conditioned on certain information obtained from its environment is shown to exhibit universal behaviours---
we investigate the entanglement properties of FQH states upon conditioning on information from projective measurements. Specifically, we study the \textit{projected ensemble} (PE), a collection of quantum states on a subsystem $\mathrm{A}$ generated by projectively measuring its complement $\mathrm{B}\equiv\bar{\mathrm{A}}$. The PE is said to be an \textit{unravelling} of the RDM, and it can be understood as probing finer features of subsystem $\mathrm{A}$ than captured by the RDM. This is because the latter is obtained by tracing out the complement $\mathrm{B}$, i.e., discarding all such measurement information, whilst the former explicitly retains correlations with $\mathrm{B}$. 
Recent advances in quantum simulators have enabled access to such finer structures by taking snapshots of quantum systems through measurements~\cite{Bakr2009QuantumGasMicroscope,Sherson2010SingleAtom,
Bernien2017ManyBody,Zhang2017DynamicalPhaseTransition, yan2026}. 
The PE is also a natural object of interest in studying measurement-induced entanglement in low complexity states~\cite{mcginley_measurement-induced_2025} and critical states~\cite{Lin_probing_2023, milekhin_observable-projected_2025, khanna_measurement-induced_2026}. 

In this work, we use the framework of PEs to perform an 
unravelling of the Li--Haldane entanglement spectrum. Rather than probing only the spectrum of the RDM, we resolve its constituent post-measurement states and characterize their distribution within the Li--Haldane entanglement subspace.
We find that this distribution is highly structured: conditioning on measurement outcomes splits the edge manifold into measurement-dependent subspaces that exhibit what we term ``rank collapse". These conditional subspaces are not arbitrary; their dimensions are governed by CFT counting with quantum numbers updated by the measurement outcome, as captured by a \textit{measurement-resolved} Li--Haldane conjecture. This reveals a previously hidden hierarchy of edge subspaces nested within a single Li--Haldane sector, which for the non-Abelian Moore--Read state even relates different parity sectors of the edge CFT. We further show that this hierarchical organisation persists under coherent deformations of the measurement basis through a ``rank rigidity'' property, and survives for realistic Coulomb-interacting ground states.

{\it Projected ensemble for FQH states.}---
Consider FQH states on the sphere, where the lowest-Landau-level states $\ket{m}$ are labelled by their angular momentum projection $m=-S,\ldots,S$, with $N_{\phi}=2S$ the number of magnetic flux quanta threading the sphere \cite{fano1986configuration}. Throughout this work, we focus on the non-abelian bosonic (fermionic) Moore-Read state at filling factor $\nu=1$ ($\nu=1/2$). This admits a Jack polynomial representation $J_{\lambda}^{-(k+1)}$ with root occupation patterns $\lambda=\{20202\ldots202\}$ ($\lambda=\{1100\ldots110011\}$) \cite{bernevig2008model, bernevig2008generalized}. An important property used throughout is that the expansion coefficients in the occupation-number basis are real.

\begin{figure*}[]
\vspace{6pt}

\centering
\subfloat[]{\includegraphics[width = 1.6in]{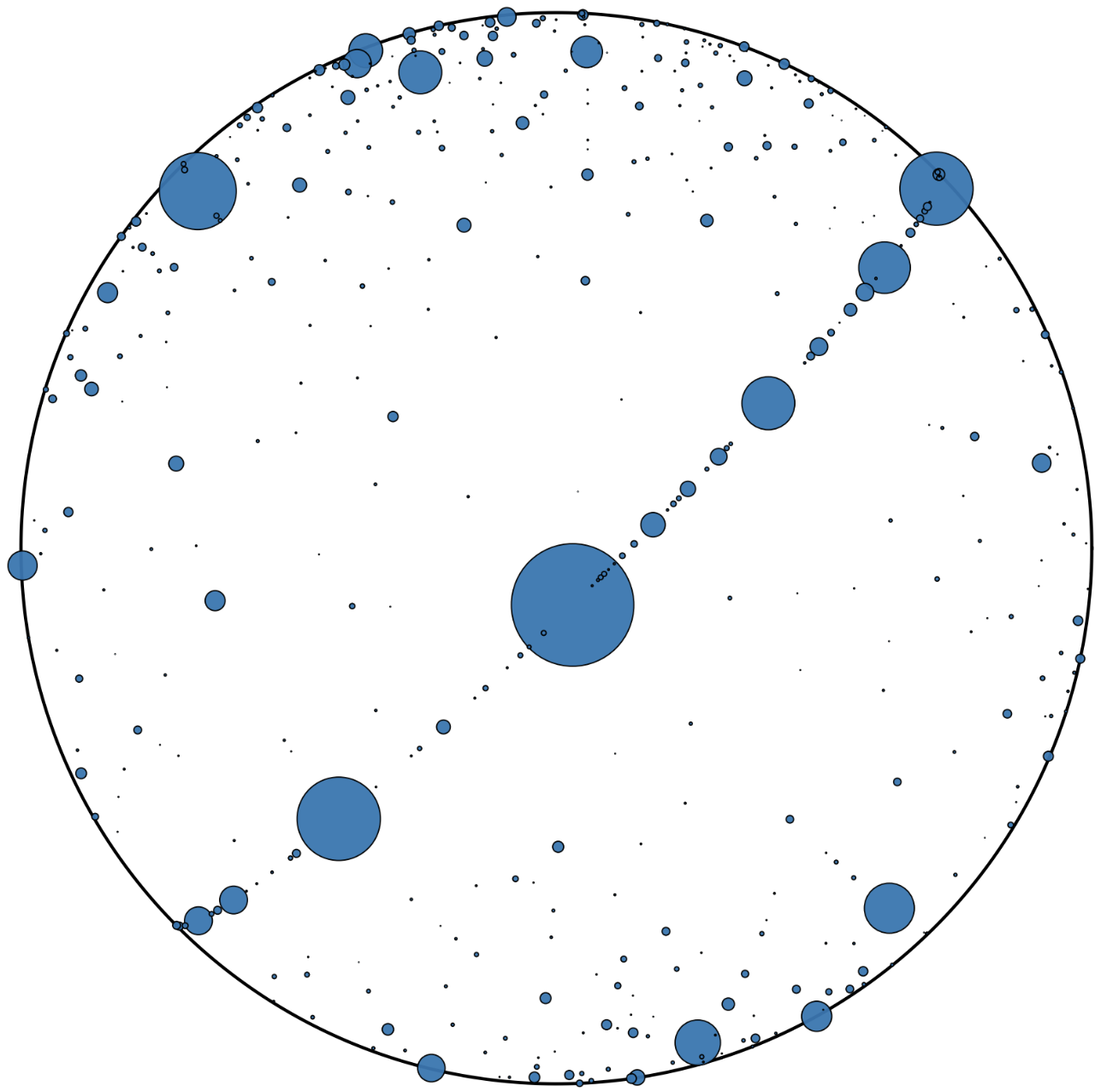}}
\hspace{0.6cm}
\subfloat[]{\includegraphics[width = 2.4in]{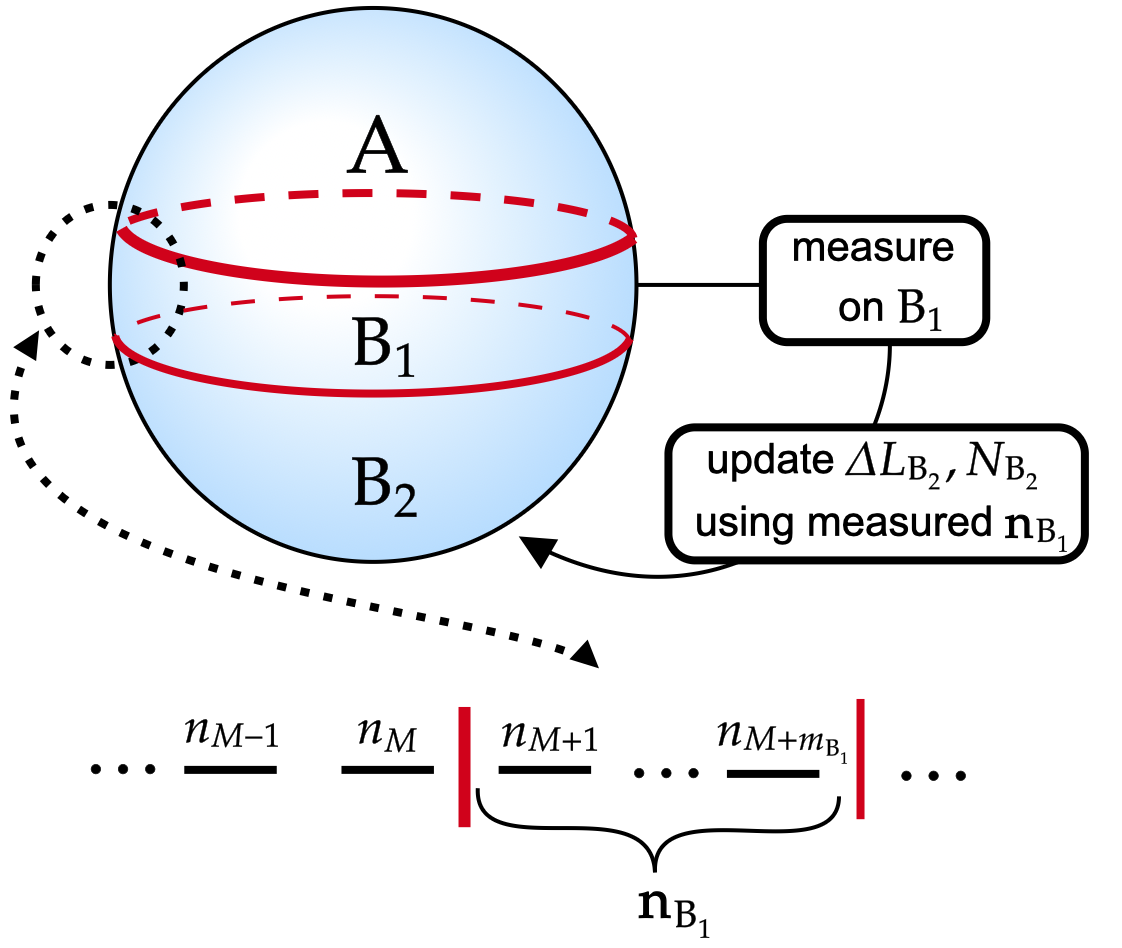}}
\hspace{0.6cm}
\subfloat[]{\includegraphics[width = 1.92in]{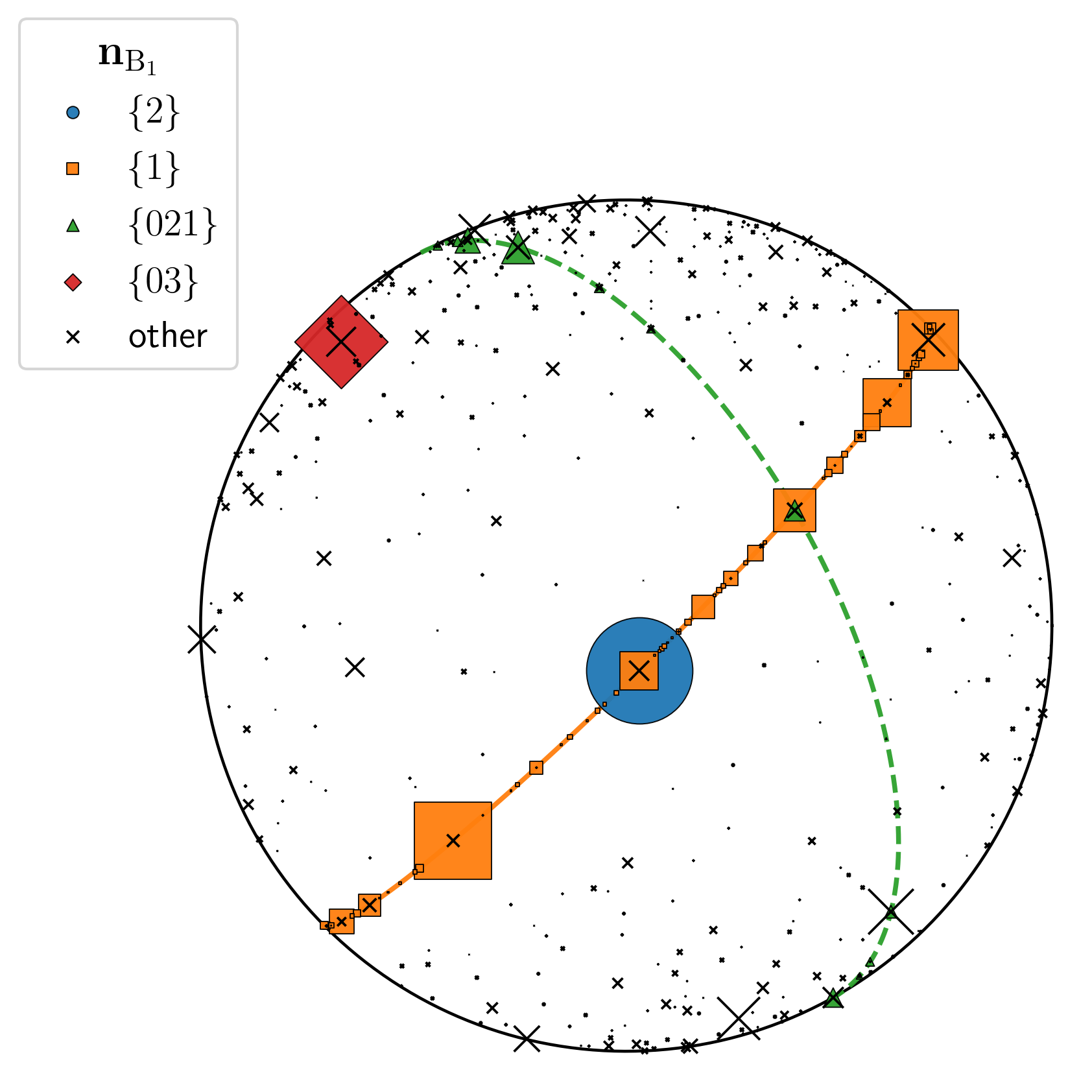}} 
\caption{
(a) Full PE in the $\Delta L_{\mathrm A}=2$ sector of the $\nu=1$, $N=20$ bosonic Moore--Read state. The states span a rank-$3$ Li--Haldane edge manifold and are represented on the disk, with antipodal identification on the boundary, corresponding to $\mathbb{RP}^{2}$. Point sizes are proportional to their PE probabilities.
(b) Schematic of partitioning the measured subsystem into $\mathrm B=\mathrm B_1\cup\mathrm B_2$. Conditioning on the occupation configuration $\mathbf n_{\mathrm B_1}$ of the orbitals nearest the entanglement cut fixes the particle number and angular momentum remaining in $\mathrm B_2$, thereby defining an effective bipartition $\mathrm A|\mathrm B_2$ and constraining the corresponding edge-state subspace of $\mathrm A$.
(c) Labelled version of a). Conditioning on
$\mathbf n_{\mathrm B_1}=\{1\}$ or $\{021\}$ produces rank collapse
$3\rightarrow2$, while $\mathbf n_{\mathrm B_1}=\{2\}$ produces
$3\rightarrow1$. In particular, odd-parity outcomes in $\mathrm B_1$ generate conditional subspaces whose dimensions follow the
$U(1)\times\psi$ CFT counting, even though the full Li--Haldane manifold is spanned by edge modes in the $U(1)\times\mathbbm{1}$ sector.
}
\label{PEDistributionsDeltaL=2}
\vspace{6pt}
\end{figure*}

Following the original construction of Li and Haldane, we consider an orbital bipartition of the FQH state (see Fig.~\ref{PEDistributionsDeltaL=2}). Introducing an angular momentum cut-off $M$ divides the single-particle orbitals into two subsystems: 
$
\mathrm{A}=\{m|m \leq M\}$, and $\mathrm{B}=\{m|m > M\}.$
The many-body state can then be written as
\begin{equation}
\ket{\Psi_{\text{FQH}}}=\sum_{\mathbf{n}_\mathrm{A},\mathbf{n}_\mathrm{B}}
c_{\mathbf{n}_\mathrm{A},\mathbf{n}_\mathrm{B}}
\ket{\mathbf{n}_\mathrm{A}}\otimes\ket{\mathbf{n}_\mathrm{B}},
\end{equation}
where $\mathbf{n}_\mathrm{A}$ and $\mathbf{n}_\mathrm{B}$ denote occupation-number configurations in subsystems $\mathrm{A}$ and $\mathrm{B}$, respectively. Upon performing a projective measurement on subsystem $\mathrm{B}$ in the occupation-number basis of orbitals $m$, the quantum state on subsystem $\mathrm{A}$ collapses into a post-measurement state $|\psi_{\mathrm{A}}(\mathbf{n}_{\mathrm{B}})\rangle=\frac{1}{\sqrt{p(\mathbf{n}_{\mathrm{B}})}} \sum_{\mathbf{n}_{\mathrm{A}}}c_{\mathbf{n}_{\mathrm{A}},\mathbf{n}_{\mathrm{B}}}|\mathbf{n}_{\mathrm{A}}\rangle$, where $\mathbf{n}_\mathrm{B}$ denotes the measurement outcome, and $p(\mathbf{n}_{\mathrm{B}})=\sum_{\mathbf{n}_{\mathrm{A}}}
|c_{\mathbf{n}_{\mathrm{A}},\mathbf{n}_{\mathrm{B}}}|^2$ is the Born probability with which the outcome occurs.
The collection of post-measurement states and probabilities
defines the PE
\begin{equation}
\mathcal{E}_{\mathrm{A}} =\{p(\mathbf{n}_{\mathrm{B}}), |\psi_{\mathrm{A}}(\mathbf{n}_{\mathrm{B}})\rangle\},
\end{equation}
the object of interest in this study.
Note the RDM $\rho_{\mathrm{A}}$ can be reconstructed as the first moment of the PE, i.e.,
$\rho_{\mathrm{A}}=\sum_{\mathbf{n}_{\mathrm{B}}}
p(\mathbf{n}_{\mathrm{B}})
\ket{\psi_{\mathrm{A}}(\mathbf{n}_{\mathrm{B}})}
\bra{\psi_{\mathrm{A}}(\mathbf{n}_{\mathrm{B}})}.
$

For an orbital bipartition, the RDM is block diagonal with respect to both the particle number $N_{\mathrm{A}}$ and the angular-momentum quantum number $\Delta L_{\mathrm{A}}$, i.e., $\rho_{\mathrm{A}}=\bigoplus_{N_{\mathrm{A}},\Delta L_{\mathrm{A}}}
p(N_\mathrm{A}, \Delta L_\mathrm{A})\rho_{\mathrm{A}}(N_{\mathrm{A}},\Delta L_{\mathrm{A}})$, where $p(N_\mathrm{A}, \Delta L_\mathrm{A})$ is the probability of finding subsystem $\mathrm{A}$ in the $(N_\mathrm{A}, \Delta L_\mathrm{A})$ sector. 
Correspondingly,
since the measurement basis is compatible with 
these sectors, 
the full PE is decomposed into
$\mathcal{E}_{\mathrm{A}}(N_{\mathrm{A}},\Delta L_{\mathrm{A}})=\{p(\mathbf{n}_\mathrm{B}|N_\mathrm{A},\Delta L_\mathrm{A}), |\psi_\mathrm{A}(\mathbf{n}_\mathrm{B}|N_\mathrm{A},\Delta L_\mathrm{A})\rangle\}$, which separately unravels each $\rho_\mathrm{A}(N_\mathrm{A},\Delta L_\mathrm{A})$. 
The Li--Haldane conjecture~\cite{Li-08-EntanglementSpectrum} states that the low-lying entanglement spectrum in a given sector exhibits the same level counting as the edge CFT associated with the underlying topological phase. For model states constructed using Jack polynomials, the rank of each block of RDM is bounded by the corresponding CFT counting,
\begin{equation}
\label{LiHaldaneconjecturemodel}
\mathrm{rank}\left[\rho_{\mathrm{A}}(N_{\mathrm{A}},\Delta L_{\mathrm{A}})\right]
\le
\mathcal{N}_{\mathrm{CFT}}(\Delta L_{\mathrm{A}}, \pi_{\mathrm{A}}),
\end{equation}
where $\mathcal{N}_{\mathrm{CFT}}(\Delta L_{\mathrm{A}}, \pi_{\mathrm{A}})$ denotes the number of independent edge-CFT states in the sector $(N_{\mathrm{A}},\Delta L_{\mathrm{A}})$, and $\pi_{A}=N_{A} \mod 2$ is the parity of subsystem $\mathrm{A}$ (see Supplementary Material for further details on the counting function $\mathcal{N}_{\mathrm{CFT}}$ for different FQH states). 
The inequality (\ref{LiHaldaneconjecturemodel}) is expected to be saturated in the thermodynamic limit as a consequence of the bulk--edge correspondence \cite{chandran2011bulk}. Our central question is whether the sector-resolved PEs 
$\mathcal{E}_{\mathrm{A}}(N_{\mathrm{A}},\Delta L_{\mathrm{A}})$ contain topological information beyond that encoded in the RDMs $\rho_{\mathrm{A}}(N_\mathrm{A}, \Delta L_\mathrm{A})$. 

Before analysing the distribution of states in the PE, we note that the subspaces themselves admit a microscopic identification. Every state in the PE necessarily lies within the support of the RDM, so the rank bound implied by the Li--Haldane conjecture (counting discussed above) constrains the PE to the same low-dimensional subspace selected by the edge CFT. This subspace can be identified microscopically with projected edge modes (PEMs), constructed from the conventional $N_{A}$-particle FQH edge state projected onto the orbitals of subsystem A, i.e. $m\leq M$ \cite{yan2019bulk}. Such PEMs were recently shown to remain gapless under non-local angular momentum confining potentials, with a close connection to bulk collective excitations including the graviton and quasielectron modes \cite{spasic2025gapless}. We verify this correspondence by computing the projector overlap between the RDM eigenstates and the projected edge-mode subspace, finding near-unity overlap up to small numerical error, for sectors $\Delta L_{\mathrm{A}}=1,\ldots,8$.

{\it Numerical investigations and rank collapse}.---Having defined the PE for an FQH state and identified its basis in terms of PEMs, we now investigate its structure numerically.  
Since the Jack polynomial model states considered here have purely real coefficients in the occupation-number basis, every projected state may be represented by a normalized real vector in the corresponding CFT subspace. Physical states are defined only up to an overall sign, and therefore naturally form a real projective Hilbert space: in a sector of dimension $n$, the PE is supported on $
\mathbb{RP}^{n-1}$.
For the Moore-Read state at $\Delta L_{\mathrm{A}}=2$ and in an even parity sector, we have $n=\mathcal{N}_{\mathrm{CFT}}(\Delta L_{\mathrm{A}}=2, \pi_{\mathrm{A}}=0)=3$, and the PE is supported on $\mathbb{RP}^{2}$, which we represent on the northern hemisphere of $S^{2}$ with antipodal identification on the equator (visualized on a disk by projecting onto the $xy$ plane). Without loss of generality
and due to ease of visualisation of these subspaces, we will focus on the sector mentioned above.

In Fig.~\ref{PEDistributionsDeltaL=2} (a), we show the distribution of projected-ensemble states in the $\Delta L_{\mathrm{A}}=2$ sector for the bosonic Moore-Read state. Several notable features emerge. Firstly, rather than being distributed uniformly throughout the available state space, the projected states concentrate near specific regions of the manifold. Distinct high-probability states are visible, exhibiting extended structures that follow great-circle trajectories on $S^2$. 

The geometric features observed in Fig.~\ref{PEDistributionsDeltaL=2} (a) can be understood as manifestations of one underlying phenomenon that we term \textit{rank collapse}. Let the PE in a given sector be supported on an $n$-dimensional subspace, as determined by the Li--Haldane counting. We say that rank collapse occurs when a subset of projected states is confined to a $k$-dimensional subspace, with $k<n$. In this language, isolated high-probability points correspond to rank-$1$ subspaces, while the great-circle trajectories observed in the $n=3$ sector correspond to rank-$2$ subspaces embedded within the full three-dimensional support.

To understand the origin of rank collapse, we seek to identify which measurements determine the rank-$k$ subspaces. Since multiple measurement outcomes produce the same rank-$k$ conditional subspace, the relevant information must be encoded only in a subset of the measured degrees of freedom. Because edge modes shared between $\mathrm{A}$ and $\mathrm{B}$ are most strongly correlated near the orbital cut, it is natural to expect that the orbitals closest to the cut predominantly determine the edge degrees of freedom.

Motivated by this observation, we further partition the measured subsystem into $\mathrm{B}=\mathrm{B}_1\cup\mathrm{B}_2$, where $\mathrm{B}_1=\{m|M<m\leq M+m_{\mathrm{B}_{1}} \}$ contains the $m_{\mathrm{B}_1} $ orbitals closest to the cut and $\mathrm{B}_2=\{m|m>M+m_{\mathrm{B}_{1}} \}$ the remaining orbitals (see Fig.~\ref{PEDistributionsDeltaL=2} (b)). Conditioning only on a measurement outcome in $\mathrm{B}_1$ then leaves $\mathrm{B}_2$ unmeasured, allowing us to define the corresponding conditional reduced density matrices (CRDMs) $\rho_{\mathrm{A}|\textbf{n}_{\mathrm{B}_{1}}}=\sum_{\mathbf{n}_{\mathrm{B}_{2}}}
p(\mathbf{n}_{\mathrm{B}_{2}}|\mathbf{n}_{\mathrm{B}_{1}})
\ket{\psi_{\mathrm{A}}(\mathbf{n}_{\mathrm{B}_{1}},\mathbf{n}_{\mathrm{B}_{2}})}
\bra{\psi_{\mathrm{A}}(\mathbf{n}_{\mathrm{B}_{1}},\mathbf{n}_{\mathrm{B}_{2}})}$, obtained by conditioning on measurement outcomes in $\mathrm{B}_1$. We may then ask how the rank of this conditional ensemble depends on the outcome observed in $\mathrm{B}_1$.
The measurement outcome on $\mathrm{B}_{1}$ fixes the particle number and angular momentum of the remaining subsystem $\mathrm{B}_2$, thereby defining a new effective bipartition $\mathrm{A}|\mathrm{B}_{2}$. As the Li--Haldane counting depends only on these quantum numbers, it is natural to conjecture that the rank of the CRDM is determined entirely by the corresponding effective edge sector.

{\it Measurement-resolved Li--Haldane conjecture.}---This leads us to the following measurement-resolved version of the Li--Haldane conjecture, applied to CRDMs:
\begin{equation}
\label{rankcollapseconjecture}
\text{rank}\left[\rho_{\mathrm{A|\textbf{n}_{\mathrm{B}_{1}}}}(N_{\mathrm{A}},\Delta L_{\mathrm{A}})\right] \leq \mathcal{N}_{\text{CFT}}(\Delta L_{\mathrm{B}_{2}}(\textbf{n}_{\mathrm{B}_{1}}), \pi_{\mathrm{B}_{2}} (\mathbf{n}_{\mathrm{B}_{1}})).
\end{equation}
The function $\Delta L_{\mathrm{B}_{2}}(\mathbf{n}_{\mathrm{B}_{1}})$ can be understood intuitively as the updated angular momentum shift in $\mathrm{B}_{2}$ after conditioning on the measurement outcome $\mathrm{B}_{1}$. For example, measuring a sequence which corresponds to the root of the FQH state, or any of its squeezed descendants, would not update the angular momentum in $\mathrm{B}_{2}$. However, if we measure an angular momentum boost relative to one of these configurations, we know that the angular momentum in $\mathrm{B}_{2}$ is constrained. Formalising this intuition gives the expression $\Delta L_{\mathrm{B}_{2}}(\textbf{n}_{\mathrm{B}_{1}}) = \mathrm{min}( \Delta L_{\mathrm{B}} + L_{\text{root}}(N_{\mathrm{B}}) - L_{B_{1}}(\textbf{n}_{\mathrm{B}_{1}}) - L_{\text{root}}(N_{\mathrm{B}_{2}}),\Delta L_{\mathrm{B}})$, where $L_{\text{root}}(N)$ is the total angular momentum of the root configuration of the corresponding $N$--particle FQH state. The minimum function guarantees that $\Delta L_{\mathrm{B}_{2}}$ is upper bounded by $\Delta L_{\mathrm{B}}$, to avoid a rank counting larger than that of the original subspace of $\rho_{\mathrm{A}}(N_{\mathrm{A}},\Delta L_{\mathrm{A}})$. The updated parity is given by $\pi_{\mathrm{B}_{2}}(\mathbf{n}_{\mathrm{B}_{1}}) = N_{\mathrm{B}} - N_{\mathrm{B}_{1}} \mod 2$.  The original Li--Haldane conjecture is recovered as the special case $\mathbf{n}_{\mathrm{B}_1}=\{\}$, corresponding to no conditioning on measurement outcomes.

We have tested Eq.~(\ref{rankcollapseconjecture}) extensively in finite-size systems (see classification of subspaces in Fig.~\ref{PEDistributionsDeltaL=2} (c)) and found no violations within the range accessible to numerics. Moreover, the bound is frequently saturated e.g., for the bosonic Moore-Read with $N=18$ , saturation is observed for all measurement outcomes with $m_{\mathrm{B}_{1}}\leq 4$ in sectors $\Delta L_{\mathrm{A}}=1,2,3,4$. These observations suggest that, analogously to the original Li--Haldane conjecture, the conditional bound becomes saturated in the thermodynamic limit. Further evidence is provided by the finite-size scaling of the saturation probability presented in the Supplementary Material, which indicates that deviations from saturation become progressively rarer with increasing system size. 

Beyond confirming the conditional CFT counting, these results reveal a hidden feature of the Li--Haldane edge manifold. Conditioning on measurement outcomes does not produce arbitrary lower-dimensional subspaces. Instead, the PEMs reorganize into distinguished subspaces whose dimensions are precisely those predicted by the CFT counting of the effective bipartition. In this sense, the PE uncovers a hidden hierarchy of edge subspaces within a single Li--Haldane sector.

This hierarchy is particularly striking for the non-Abelian Moore-Read state, whose edge theory is described by $U(1)\times \mathrm{Ising}$. The Li--Haldane edge manifold associated with a fixed particle-number parity, for example the even-parity $U(1)\times \mathbbm{1}$ sector, contains conditional subspaces whose dimensions coincide with the CFT counting of the opposite-parity $U(1)\times \psi$ sector. While the conditional states remain PEMs of the original Li--Haldane sector, their internal organization is governed by the CFT of a different sector. Thus, although the conditional states themselves remain PEMs spanning the original Li--Haldane manifold, they organize into nested subspaces whose dimensions are dictated by the CFT counting of a different topological sector. The PE therefore reveals that a single Li--Haldane edge manifold already encodes a hidden hierarchy of CFT sectors, exposing considerably richer structure than is apparent from its total dimension alone.

{\it Rank rigidity.}---A natural question is whether the rank-collapse structure depends on the choice of measurement basis. So far, we have considered projective measurements in the occupation-number basis of subsystem $\mathrm{B}_{1}$. We find that conditional CFT counting extends naturally to arbitrary linear combinations of occupation-number configurations with the same particle number and angular momentum, thus generalising Eq.~(\ref{rankcollapseconjecture}) to any measurement basis within a fixed symmetry sector. We refer to the preservation of the rank-collapse structure under continuous interpolation between such measurement bases as \textit{rank rigidity}. Consequently, rank collapse is not a property of individual occupation-number configurations but of the measurement sector itself.
 
Rank rigidity imposes strong constraints on the geometry of the PE. The conditioned subspaces associated with different measurement outcomes cannot vary independently, but must deform coherently so as to preserve their rank under basis rotations. As a result, the probability distribution of projected states within the PE is highly constrained. In the Supplementary Material, we show that these constraints are sufficiently restrictive to derive analytic mappings between rank-collapsed subspaces. These results provide a first characterization of the internal geometry of the PE and motivate further study of its underlying structure.

\begin{figure}[t]
\subfloat[]{\includegraphics[width = 1.75in]{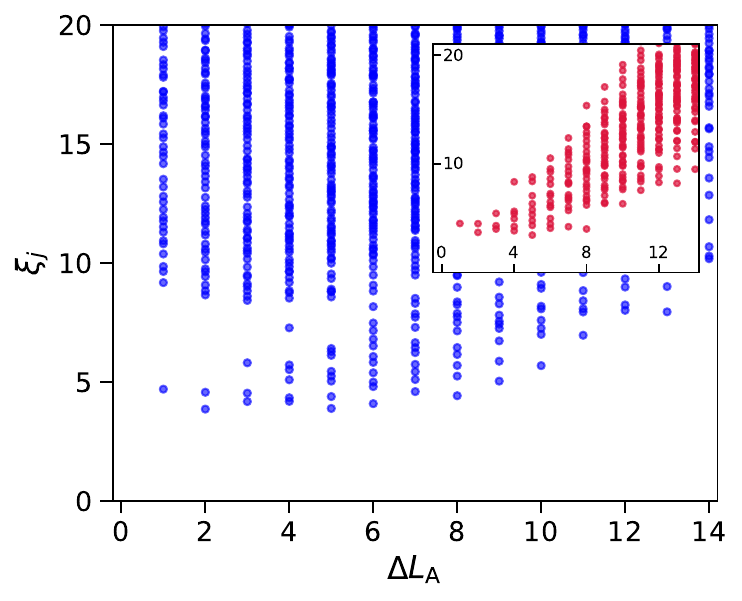}}
\subfloat[]{\includegraphics[width = 1.75in]{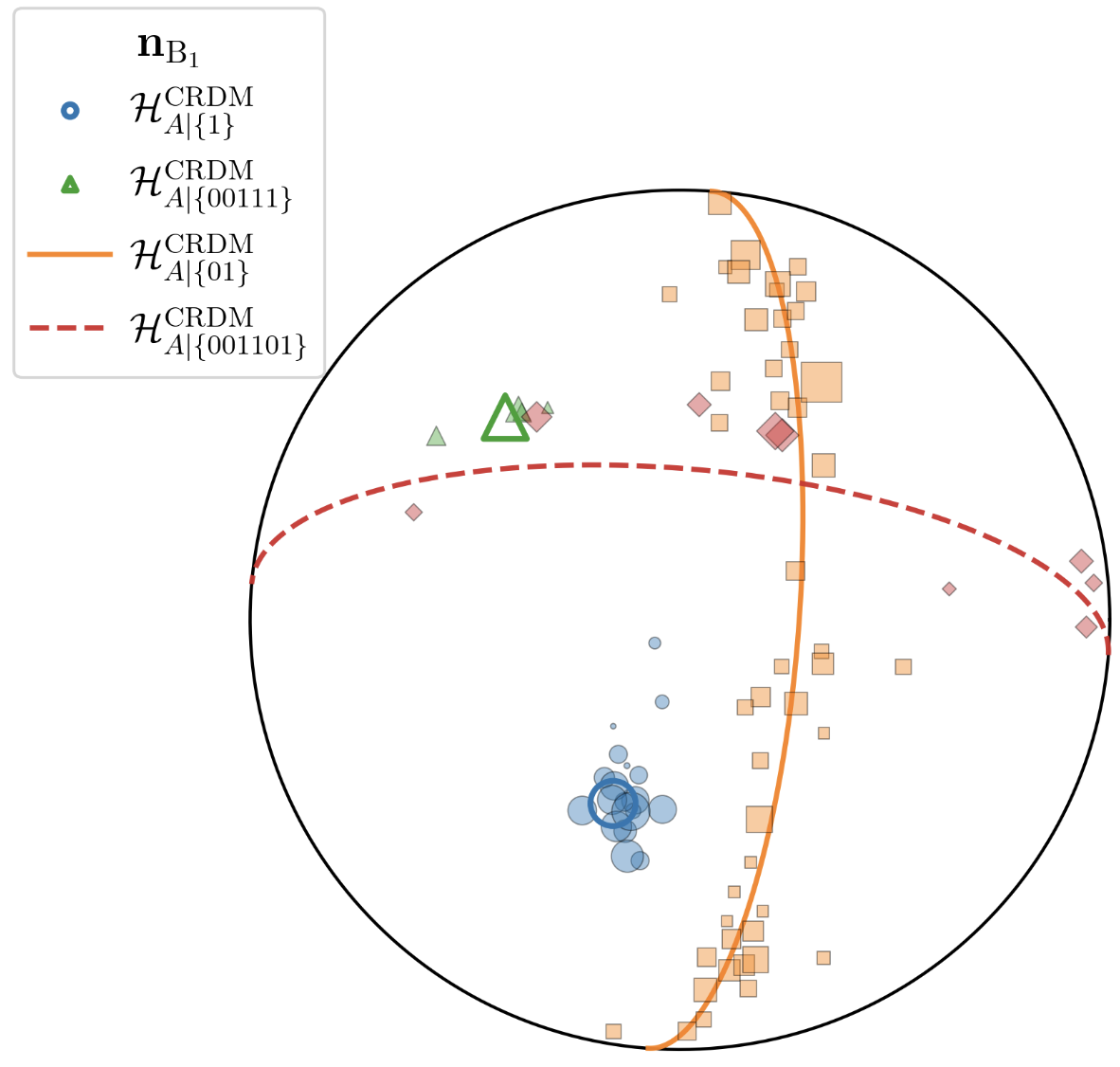}}\\
\caption{
(a) CRDM entanglement spectrum of the $\nu=5/2$, $N=16$ Coulomb ground state, for
$\mathbf{n}_{\mathrm{B}_1}=\{1\}$. The low-lying sector exhibits the CFT
counting predicted by Eq.~(\ref{rankcollapseconjecture}) and is separated
from the non-universal UV sector by an entanglement gap, analogous to the
Li--Haldane structure of the full RDM. The corresponding low-lying
eigenstates span the conditional edge subspace
$\mathcal{H}_{A|\mathbf n_{\mathrm B_1}}^{\mathrm{CRDM}}$. (The inset is the corresponding entanglement spectrum of the fermionic $\nu=1/2$ Moore-Read state)
(b) PE of the $\nu=5/2$ Coulomb ground state, with the interaction projected to the second Landau level. The ensemble is projected onto the universal edge subspace of the full Coulomb RDM, and the corresponding conditional subspaces $\mathcal{H}_{A|\mathbf n_{\mathrm B_1}}^{\mathrm{CRDM}}$ are indicated. These subspaces occupy essentially the same positions within the edge manifold as in the Moore-Read model state, while the full Coulomb PE exhibits a small broadening around the ideal rank-collapse manifolds.}
\label{CoulombFigure}
\end{figure}

{\it Realistic interactions.}---Finally, we ask whether the measurement-resolved structure found for model
wavefunctions survives realistic interactions. We consider the Coulomb ground
state at half filling of the second Landau level, corresponding to the $\nu=5/2$ FQH state. As shown in Fig.~\ref{CoulombFigure}(a), the CRDM entanglement spectrum, $\xi_j=-\log\lambda_j$, exhibits a low-lying universal sector separated from the non-universal UV spectrum by an entanglement gap, closely analogous to the conventional Li--Haldane spectrum. The number of states below this gap follows the same conditional CFT counting as for the Moore-Read model state. We therefore define the corresponding conditional edge subspace as
\begin{equation}
\mathcal{H}_{A|\mathbf n_{\mathrm B_1}}^{\mathrm{CRDM}} = \mathrm{span} \left\{ \ket{\lambda_j^{\mathrm{CRDM}}(\mathbf n_{\mathrm B_1})}\right\}_{j=1}^{k},
\end{equation} where $\ket{\lambda_j^{\mathrm{CRDM}}(\mathbf n_{\mathrm B_1})}$ are the lowest energy eigenstates of the CRDM, whose dimension is fixed by Eq.~(\ref{rankcollapseconjecture}). The states
spanning $\mathcal{H}_{A|\mathbf n_{\mathrm B_1}}^{\mathrm{CRDM}}$ have near-unit support within the universal edge manifold of the full Coulomb RDM, showing that the conditional edge structure identified for the model state persists for realistic interactions.

Projecting the Coulomb PE onto this RDM edge manifold reveals the same measurement-resolved organization as in the fermionic Moore-Read state [Fig.~\ref{CoulombFigure}(b)]. The conditional subspaces $\mathcal{H}_{A|\mathbf n_{\mathrm B_1}}^{\mathrm{CRDM}}$ occupy essentially
the same positions within the edge manifold as their model-state counterparts, while the individual post-measurement Coulomb states cluster around them. The small broadening away from the ideal rank-collapse manifolds reflects residual UV admixture in the Coulomb post-measurement states; we analyze its origin in detail in the Supplementary Material. Thus, the Coulomb state reproduces not only the universal Li--Haldane counting but also the measurement-resolved internal organization of the edge manifold.

{\it Discussion and outlook.}---
In this work, we studied universal features of the Li--Haldane entanglement subspace of FQH states through the lens of the projected ensemble, the collection of conditional quantum states on a subsystem resolved by measurements of its complement. We uncovered a richer internal structure than that accessible from the RDM alone. This led us to formulate a measurement-resolved version of the Li--Haldane conjecture, relating conditional CFT counting to projective measurements on subsets of orbitals. We applied this conjecture to Moore-Read states and showed that the conditional subspaces are spanned by PEMs of one CFT sector, even though they may be constrained to subspaces of a different CFT sector. We further demonstrated that these measurement-resolved features persist for realistic Coulomb-interacting ground states. 

Our results show that the PE provides an operational refinement of the entanglement spectrum, revealing universal structure hidden by averaging over measurement outcomes. More broadly, they show that measurement-resolved entanglement offers a useful new perspective on the interplay between quantum measurements, edge CFTs, and topological order.

\textit{Acknowledgements -- } 
This work was supported by the EPSRC [grant number EP/V062654/1], a Simons Investigator Award [Grant No. 511029] and a Cambridge International Scholarship provided by the Cambridge Trust. 
W.~W.~H.~is supported by the National Research Foundation (NRF), Singapore, through the NRF Fellowship NRF-NRFF15-2023-0008, and through the National Quantum Office, hosted in A*STAR, under its Centre for Quantum Technologies Funding Initiative (S24Q2d0009). The initial idea for this work originated from discussions at the ICTP Trieste summer school.
For the purpose of open access, the authors have applied a creative commons attribution (CC BY) licence to any author accepted manuscript version arising.

\nocite{*}
\bibliography{library.bib}

@article{Li-08-EntanglementSpectrum,
  title = {{Entanglement Spectrum as a Generalization of Entanglement Entropy: Identification of Topological Order in Non-Abelian Fractional Quantum Hall Effect States}},
  author = {Li, Hui and Haldane, F. D. M.},
  journal = {Phys. Rev. Lett.},
  volume = {101},
  issue = {1},
  pages = {010504},
  numpages = {4},
  year = {2008},
  month = {Jul},
  publisher = {American Physical Society},
  doi = {10.1103/PhysRevLett.101.010504},
  url = {https://link.aps.org/doi/10.1103/PhysRevLett.101.010504}
}

@article{laflorencie2016quantum,
  title={Quantum entanglement in condensed matter systems},
  author={Laflorencie, Nicolas},
  journal={Physics Reports},
  volume={646},
  pages={1--59},
  year={2016},
  publisher={Elsevier}
}

@article{dubail2011real,
  title={{Real-space entanglement spectrum of quantum Hall systems}},
  author={Dubail, Jerome and Read, N and Rezayi, EH},
  journal={arXiv preprint arXiv:1111.2811},
  year={2011}
}

@article{sterdyniak2011extracting,
  title={Extracting excitations from model state entanglement},
  author={Sterdyniak, A and Regnault, N and Bernevig, B Andrei},
  journal={Phys. Rev. Lett.},
  volume={106},
  number={10},
  pages={100405},
  year={2011},
  publisher={APS}
}

@article{qi2012general,
  title={General Relationship between the Entanglement Spectrum and the Edge State Spectrum of Topological Quantum States},
  author={Qi, Xiao-Liang and Katsura, Hosho and Ludwig, Andreas WW},
  journal={Phys. Rev. Lett.},
  volume={108},
  number={19},
  pages={196402},
  year={2012},
  publisher={APS}
}

@article{levin2006detecting,
  title={Detecting topological order in a ground state wave function},
  author={Levin, Michael and Wen, Xiao-Gang},
  journal={Phys. Rev. Lett.},
  volume={96},
  number={11},
  pages={110405},
  year={2006},
  publisher={APS}
}

@article{kitaev2006topological,
  title={Topological entanglement entropy},
  author={Kitaev, Alexei and Preskill, John},
  journal={Phys. Rev. Lett.},
  volume={96},
  number={11},
  pages={110404},
  year={2006},
  publisher={APS}
}

@article{wen1995topological,
  title={{Topological orders and edge excitations in fractional quantum Hall states}},
  author={Wen, Xiao-Gang},
  journal={Advances in Physics},
  volume={44},
  number={5},
  pages={405--473},
  year={1995},
  publisher={Taylor \& Francis}
}

@article{milovanovic1996edge,
  title={{Edge excitations of paired fractional quantum Hall states}},
  author={Milovanovi{\'c}, M and Read, Nicholas},
  journal={Physical Review B},
  volume={53},
  number={20},
  pages={13559},
  year={1996},
  publisher={APS}
}

@article{moore1991nonabelions,
  title={{Nonabelions in the fractional quantum Hall effect}},
  author={Moore, Gregory and Read, Nicholas},
  journal={Nuclear Physics B},
  volume={360},
  number={2-3},
  pages={362--396},
  year={1991},
  publisher={Elsevier}
}

@article{fano1986configuration,
  title={{Configuration-interaction calculations on the fractional quantum Hall effect}},
  author={Fano, G and Ortolani, F and Colombo, E},
  journal={Physical Review B},
  volume={34},
  number={4},
  pages={2670},
  year={1986},
  publisher={APS}
}

@article{bernevig2008model,
  title={{Model fractional quantum Hall states and Jack polynomials}},
  author={Bernevig, B Andrei and Haldane, FDM},
  journal={Phys. Rev. Lett.},
  volume={100},
  number={24},
  pages={246802},
  year={2008},
  publisher={APS}
}

@article{bernevig2008generalized,
  title={{Generalized clustering conditions of Jack polynomials at negative Jack parameter $\alpha$}},
  author={Bernevig, B Andrei and Haldane, FDM},
  journal={Physical Review B—Condensed Matter and Materials Physics},
  volume={77},
  number={18},
  pages={184502},
  year={2008},
  publisher={APS}
}

@article{chandran2011bulk,
  title={Bulk-edge correspondence in entanglement spectra},
  author={Chandran, Anushya and Hermanns, M and Regnault, N and Bernevig, B Andrei},
  journal={Physical Review B - Condensed Matter and Materials Physics},
  volume={84},
  number={20},
  pages={205136},
  year={2011},
  publisher={APS}
}

@article{spasic2025gapless,
  title = {{Gapless edge gravitons and quasiparticles in fractional quantum Hall systems with nonlocal confinement}},
  author = {Spasic-Mlacak, Daniel and Cooper, Nigel R.},
  journal = {Phys. Rev. Res.},
  volume = {8},
  issue = {2},
  pages = {L022063},
  numpages = {8},
  year = {2026},
  month = {Jun},
  publisher = {American Physical Society},
  doi = {10.1103/2m3h-7x3r},
  url = {https://link.aps.org/doi/10.1103/2m3h-7x3r}
}

@article{yan2019bulk,
  title={Bulk-edge correspondence in fractional quantum hall states},
  author={Yan, Bin and Biswas, Rudro R and Greene, Chris H},
  journal={Physical Review B},
  volume={99},
  number={3},
  pages={035153},
  year={2019},
  publisher={APS}
}

@article{
yan2026,
author = {Zhiguang Yan  and Zi-Yong Ge  and Rui Li  and Yu-Ran Zhang  and Franco Nori  and Yasunobu Nakamura },
title = {Characterizing many-body dynamics with projected ensembles on a superconducting quantum processor},
journal = {Science Advances},
volume = {12},
number = {13},
pages = {eaeb8213},
year = {2026},
doi = {10.1126/sciadv.aeb8213},
URL = {https://www.science.org/doi/abs/10.1126/sciadv.aeb8213},
eprint = {https://www.science.org/doi/pdf/10.1126/sciadv.aeb8213}}

@article{ho2022exact,
  title={Exact emergent quantum state designs from quantum chaotic dynamics},
  author={Ho, Wen Wei and Choi, Soonwon},
  journal={Phys. Rev. Lett.},
  volume={128},
  number={6},
  pages={060601},
  year={2022},
  publisher={APS}
}

@article{cotler_emergent_2023,
  title = {Emergent Quantum State Designs from Individual Many-Body Wave Functions},
  author = {Cotler, Jordan S. and Mark, Daniel K. and Huang, Hsin-Yuan and Hern\'andez, Felipe and Choi, Joonhee and Shaw, Adam L. and Endres, Manuel and Choi, Soonwon},
  journal = {PRX Quantum},
  volume = {4},
  issue = {1},
  pages = {010311},
  numpages = {29},
  year = {2023},
  month = {Jan},
  publisher = {American Physical Society},
  doi = {10.1103/PRXQuantum.4.010311},
  url = {https://link.aps.org/doi/10.1103/PRXQuantum.4.010311}
}

@article{mark_maximum_2024,
	title = {Maximum {Entropy} {Principle} in {Deep} {Thermalization} and in {Hilbert}-{Space} {Ergodicity}},
	volume = {14},
	issn = {2160-3308},
	url = {https://link.aps.org/doi/10.1103/PhysRevX.14.041051},
	doi = {10.1103/PhysRevX.14.041051},
	
	number = {4},
	urldate = {2025-10-02},
	journal = {Physical Review X},
	author = {Mark, Daniel K. and Surace, Federica and Elben, Andreas and Shaw, Adam L. and Choi, Joonhee and Refael, Gil and Endres, Manuel and Choi, Soonwon},
	month = nov,
	year = {2024},
	pages = {041051},
}

@article{liu_cv_2024,
  title = {Deep Thermalization in Gaussian Continuous-Variable Quantum Systems},
  author = {Liu, Chang and Huang, Qi Camm and Ho, Wen Wei},
  journal = {Phys. Rev. Lett.},
  volume = {133},
  issue = {26},
  pages = {260401},
  numpages = {6},
  year = {2024},
  month = {Dec},
  publisher = {American Physical Society},
  doi = {10.1103/PhysRevLett.133.260401},
  url = {https://link.aps.org/doi/10.1103/PhysRevLett.133.260401}
}

@article{chang_deep_2025,
	title = {Deep {Thermalization} under {Charge}-{Conserving} {Quantum} {Dynamics}},
	volume = {6},
	issn = {2691-3399},
	url = {https://link.aps.org/doi/10.1103/PRXQuantum.6.020343},
	doi = {10.1103/PRXQuantum.6.020343},
	
	number = {2},
	urldate = {2025-10-02},
	journal = {PRX Quantum},
	author = {Chang, Rui-An and Shrotriya, Harshank and Ho, Wen Wei and Ippoliti, Matteo},
	month = jun,
	year = {2025},
	pages = {020343},

    
}

@article{liu_coherence-induced_2025,
  title = {Coherence-Induced Deep Thermalization Transition in Random Permutation Quantum Dynamics},
  author = {Liu, Chang and Ippoliti, Matteo and Ho, Wen Wei},
  journal = {Phys. Rev. Lett.},
  volume = {136},
  issue = {10},
  pages = {100404},
  numpages = {8},
  year = {2026},
  month = {Mar},
  publisher = {American Physical Society},
  doi = {10.1103/mjgs-y2z9},
  url = {https://link.aps.org/doi/10.1103/mjgs-y2z9}
}

@misc{mok_nature_2026,
      title={Nature is stingy: Universality of Scrooge ensembles in quantum many-body systems}, 
      author={Wai-Keong Mok and Tobias Haug and Wen Wei Ho and John Preskill},
      year={2026},
      eprint={2601.00266},
      archivePrefix={arXiv},
      primaryClass={quant-ph},
      url={https://arxiv.org/abs/2601.00266}, 
}

@article{mcginley2025scrooge,
  title={The Scrooge ensemble in many-body quantum systems},
  author={McGinley, Max and Schuster, Thomas},
  journal={arXiv preprint arXiv:2511.17172},
  year={2025}
}

@article{lucas_fermions_2023,
  title = {Generalized deep thermalization for free fermions},
  author = {Lucas, Maxime and Piroli, Lorenzo and De Nardis, Jacopo and De Luca, Andrea},
  journal = {Phys. Rev. A},
  volume = {107},
  issue = {3},
  pages = {032215},
  numpages = {12},
  year = {2023},
  month = {Mar},
  publisher = {American Physical Society},
  doi = {10.1103/PhysRevA.107.032215},
  url = {https://link.aps.org/doi/10.1103/PhysRevA.107.032215}
}

@article{bejan_matchgate_2025,
  title = {Matchgate Circuits Deeply Thermalize},
  author = {Bejan, Mircea and B\'eri, Benjamin and McGinley, Max},
  journal = {Phys. Rev. Lett.},
  volume = {135},
  issue = {2},
  pages = {020401},
  numpages = {7},
  year = {2025},
  month = {Jul},
  publisher = {American Physical Society},
  doi = {10.1103/v8kp-39ry},
  url = {https://link.aps.org/doi/10.1103/v8kp-39ry}
}

@article{mcginley_measurement-induced_2025,
  title = {Measurement-Induced Entanglement and Complexity in Random Constant-Depth 2D Quantum Circuits},
  author = {McGinley, Max and Ho, Wen Wei and Malz, Daniel},
  journal = {Phys. Rev. X},
  volume = {15},
  issue = {2},
  pages = {021059},
  numpages = {35},
  year = {2025},
  month = {May},
  publisher = {American Physical Society},
  doi = {10.1103/PhysRevX.15.021059},
  url = {https://link.aps.org/doi/10.1103/PhysRevX.15.021059}
}

@article{khanna_measurement-induced_2026,
  title = {Measurement-Induced Entanglement in Conformal Field Theory},
  author = {Khanna, Kabir and Vasseur, Romain},
  journal = {Phys. Rev. Lett.},
  volume = {136},
  issue = {16},
  pages = {160402},
  numpages = {8},
  year = {2026},
  month = {Apr},
  publisher = {American Physical Society},
  doi = {10.1103/b7sb-nhjq},
  url = {https://link.aps.org/doi/10.1103/b7sb-nhjq}
}

@article{milekhin_observable-projected_2025,
  doi = {10.22331/q-2025-10-20-1888},
  url = {https://doi.org/10.22331/q-2025-10-20-1888},
  title = {Observable-projected ensembles},
  author = {Milekhin, Alexey and Murciano, Sara},
  journal = {{Quantum}},
  issn = {2521-327X},
  publisher = {{Verein zur F{\"{o}}rderung des Open Access Publizierens in den Quantenwissenschaften}},
  volume = {9},
  pages = {1888},
  month = oct,
  year = {2025}
}

@article{Bakr2009QuantumGasMicroscope,
    author = {Bakr, Waseem S. and Gillen, Jonathon I. and Peng, Amy
              and F{\"o}lling, Simon and Greiner, Markus},
    title = {{A quantum gas microscope for detecting single atoms
             in a Hubbard-regime optical lattice}},
    journal = {Nature},
    volume = {462},
    pages = {74--77},
    year = {2009},
    doi = {10.1038/nature08482}
}

@article{Sherson2010SingleAtom,
    author = {Sherson, Jacob F. and Weitenberg, Christof and Endres, Manuel
              and Cheneau, Marc and Bloch, Immanuel and Kuhr, Stefan},
    title = {Single-atom-resolved fluorescence imaging of an atomic
             {Mott} insulator},
    journal = {Nature},
    volume = {467},
    pages = {68--72},
    year = {2010},
    doi = {10.1038/nature09378}
}

@article{Bernien2017ManyBody,
    author = {Bernien, Hannes and Schwartz, Sylvain and Keesling, Alexander
              and Levine, Harry and Omran, Ahmed and Pichler, Hannes
              and Choi, Soonwon and Zibrov, Alexander S. and Endres, Manuel
              and Greiner, Markus and Vuleti{\'c}, Vladan and Lukin, Mikhail D.},
    title = {Probing many-body dynamics on a 51-atom quantum simulator},
    journal = {Nature},
    volume = {551},
    pages = {579--584},
    year = {2017},
    doi = {10.1038/nature24622}
}

@article{Zhang2017DynamicalPhaseTransition,
    author = {Zhang, J. and Pagano, G. and Hess, P. W.
              and Kyprianidis, A. and Becker, P. and Kaplan, H.
              and Gorshkov, A. V. and Gong, Z.-X. and Monroe, C.},
    title = {Observation of a many-body dynamical phase transition
             with a 53-qubit quantum simulator},
    journal = {Nature},
    volume = {551},
    pages = {601--604},
    year = {2017},
    doi = {10.1038/nature24654}
}

@article{Lin_probing_2023,
  doi = {10.22331/q-2023-02-02-910},
  url = {https://doi.org/10.22331/q-2023-02-02-910},
  title = {Probing sign structure using measurement-induced entanglement},
  author = {Lin, Cheng-Ju and Ye, Weicheng and Zou, Yijian and Sang, Shengqi and Hsieh, Timothy H.},
  journal = {{Quantum}},
  issn = {2521-327X},
  publisher = {{Verein zur F{\"{o}}rderung des Open Access Publizierens in den Quantenwissenschaften}},
  volume = {7},
  pages = {910},
  month = feb,
  year = {2023}
}

@article{Holevo1973,
  author  = {Holevo, A. S.},
  title   = {Bounds for the Quantity of Information Transmitted by a Quantum Communication Channel},
  journal = {Problems of Information Transmission},
  volume  = {9},
  pages   = {177--183},
  year    = {1973}
}

\section{Supplementary Material}

\subsection{Edge CFT Counting of FQH States}

In this Appendix, we briefly summarize the conformal field theory (CFT) counting used throughout the main text. The Li--Haldane conjecture states that the support of the reduced density matrix (RDM) in each angular momentum sector is determined by the counting of edge excitations of the corresponding chiral edge conformal field theory \cite{Li-08-EntanglementSpectrum}. These counting sequences arise from the descendant structure of the chiral CFT describing the physical edge excitations of the fractional quantum Hall state \cite{wen1995topological, moore1991nonabelions, milovanovic1996edge}. 

For a subsystem containing $N_{\mathrm A}$ particles, edge excitations are labelled by the excess angular momentum
\begin{equation}
\Delta L_{\mathrm A}
=
L_{\mathrm A}-L_{\mathrm A,\text{root}},
\end{equation}
where $L_{\mathrm A,\text{root}}$ is the angular momentum of the corresponding root configuration. The CFT counting gives the number of independent edge excitations at each value of $\Delta L_{\mathrm A}$ and, according to the Li--Haldane conjecture, the dimension of the universal low-energy sector of the orbital entanglement spectrum.

\subsubsection{Laughlin state}

The edge theory of the Laughlin state is described by a single compact chiral $U(1)$ boson whose excitations are generated by bosonic density modes obeying a chiral Kac--Moody algebra \cite{wen1995topological}. The corresponding edge-state counting is given by the number of descendants of the identity representation, which is equivalent to the integer partition numbers,
\begin{equation}
\mathcal{N}_{\mathrm{CFT,Laughlin}}(\Delta L)=p(\Delta L),
\end{equation}
where $p(\Delta L)$ denotes the number of integer partitions of $\Delta L$. The first few values are listed in Table~\ref{tab:LaughlinCounting}.

\begin{table}[h]
\centering
\large
\setlength{\tabcolsep}{5pt}
\renewcommand{\arraystretch}{1.15}
\begin{tabular}{|c|ccccccccc|}
\hline
$\Delta L_{\mathrm A}$ & $0$ & $1$ & $2$ & $3$ & $4$ & $5$ & $6$ & $7$ & $8$  \\
\hline
$U(1)$ & 1 & 1 & 2 & 3 & 5 & 7 & 11 & 15 & 22 \\
\hline
\end{tabular}
\caption{Edge-CFT counting for the Laughlin state.}
\label{tab:LaughlinCounting}
\end{table}

\subsubsection{Moore--Read state}

The Moore-Read state is described by the product conformal field theory $
U(1)\times\mathrm{Ising}$,
consisting of a charged chiral boson and a neutral Majorana fermion \cite{moore1991nonabelions}. The corresponding edge-state counting is obtained from the descendant structure of the $U(1)$ and Ising conformal field theories \cite{milovanovic1996edge}. Unlike the Laughlin state, the Moore-Read edge possesses two topological sectors corresponding to the identity and Majorana primary fields of the Ising CFT. Consequently, the counting depends on the parity of the subsystem particle number, $\pi_{\mathrm A}=N_{\mathrm A}\;(\mathrm{mod}\;2)$.
We denote the corresponding counting function by
$
\mathcal{N}_{\mathrm{CFT,Moore-Read}}(\Delta L_{\mathrm A},\pi_{\mathrm A})$, where 
$\pi_{\mathrm A}=0$ ($\pi_{\mathrm A}=1$) corresponds to $U(1)\times\mathbbm{1}$ ($
U(1)\times\psi$). 
The first few values are listed in Table~\ref{tab:MRCounting}.

\begin{table}[h]
\centering
\large
\setlength{\tabcolsep}{5pt}
\renewcommand{\arraystretch}{1.15}
\begin{tabular}{|c|cccccccc|}
\hline
$\Delta L_{\mathrm A}$ & $0$ & $1$ & $2$ & $3$ & $4$ & $5$ & $6$ & $7$  \\
\hline
$U(1)\times\mathbbm{1}$ ($\pi_{\mathrm A}=0$)
& 1 & 1 & 3 & 5 & 10 & 16 & 28 & 43 \\
\hline
$U(1)\times\psi$ ($\pi_{\mathrm A}=1$)
& 1 & 2 & 4 & 7 & 13 & 21 & 35 & 55  \\
\hline
\end{tabular}
\caption{Edge-CFT counting for the Moore-Read state.}
\label{tab:MRCounting}
\end{table}

\subsection{Saturation of Rank Collapse}
\label{sec:saturation-rank-collapse}

\begin{figure}[htp]
\centering
\subfloat[$N=8$]{
\includegraphics[width=1.70in]{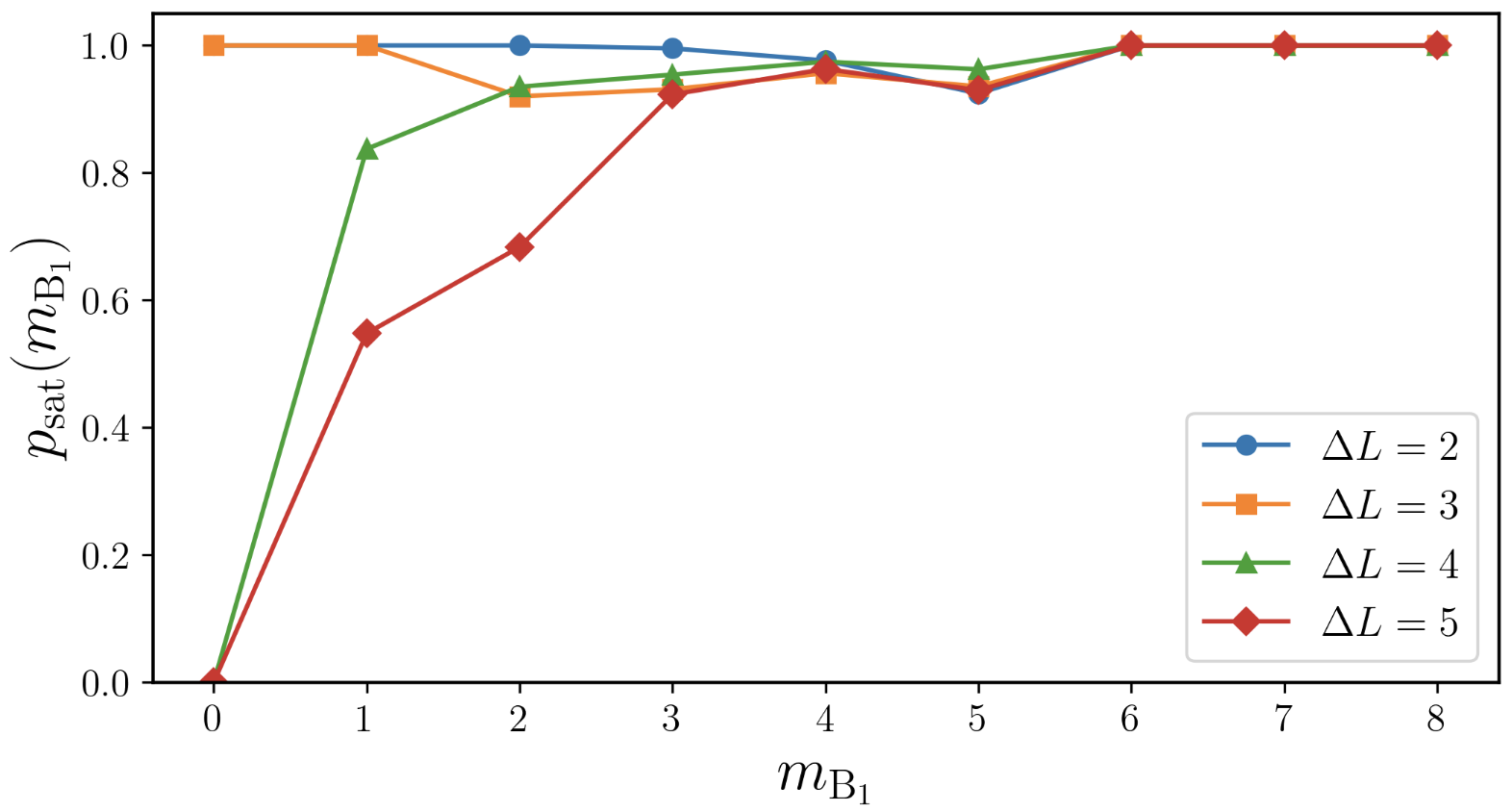}}
\subfloat[$N=10$]{
\includegraphics[width=1.70in]{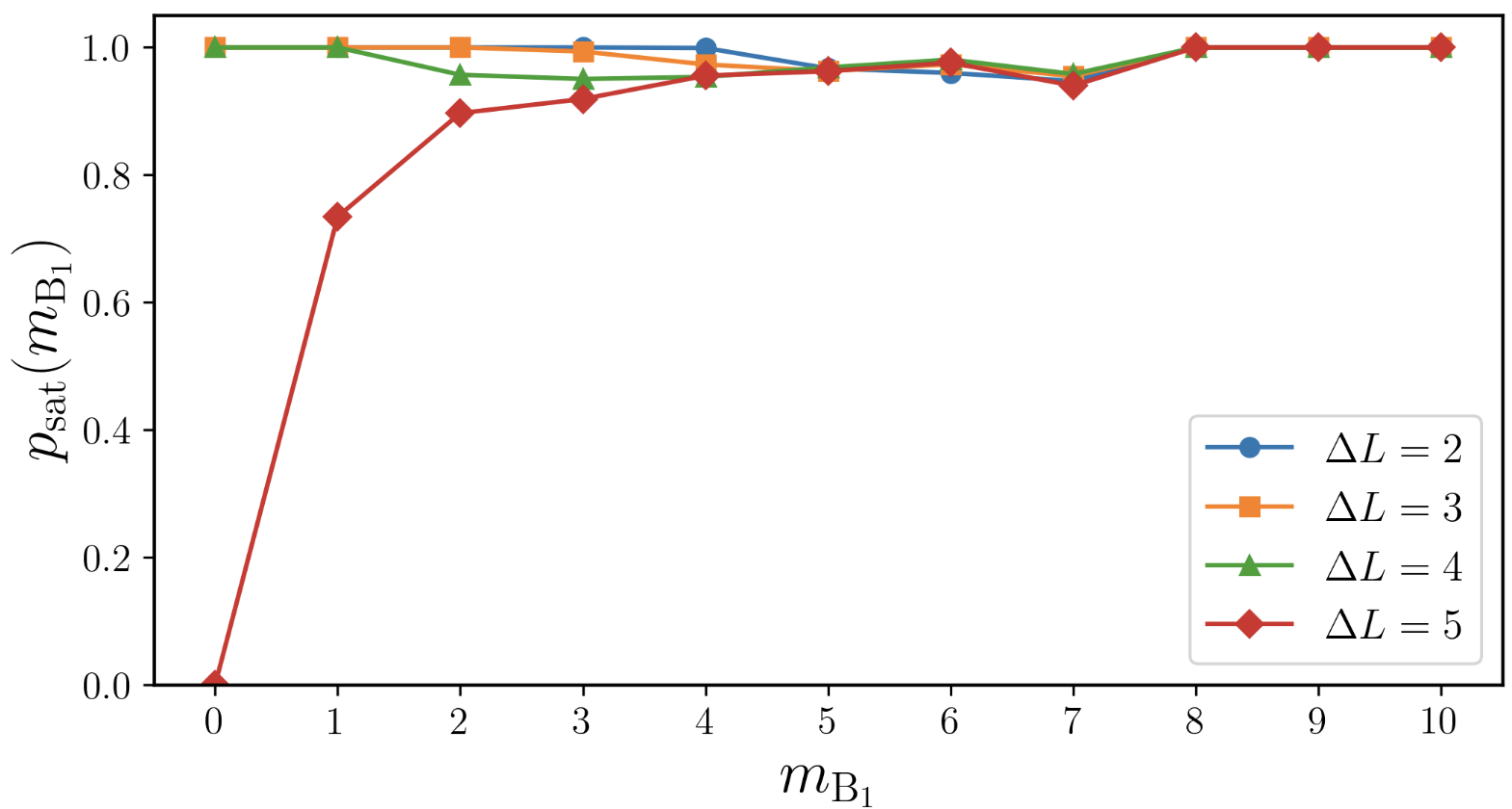}}
\\
\subfloat[$N=12$]{
\includegraphics[width=1.70in]{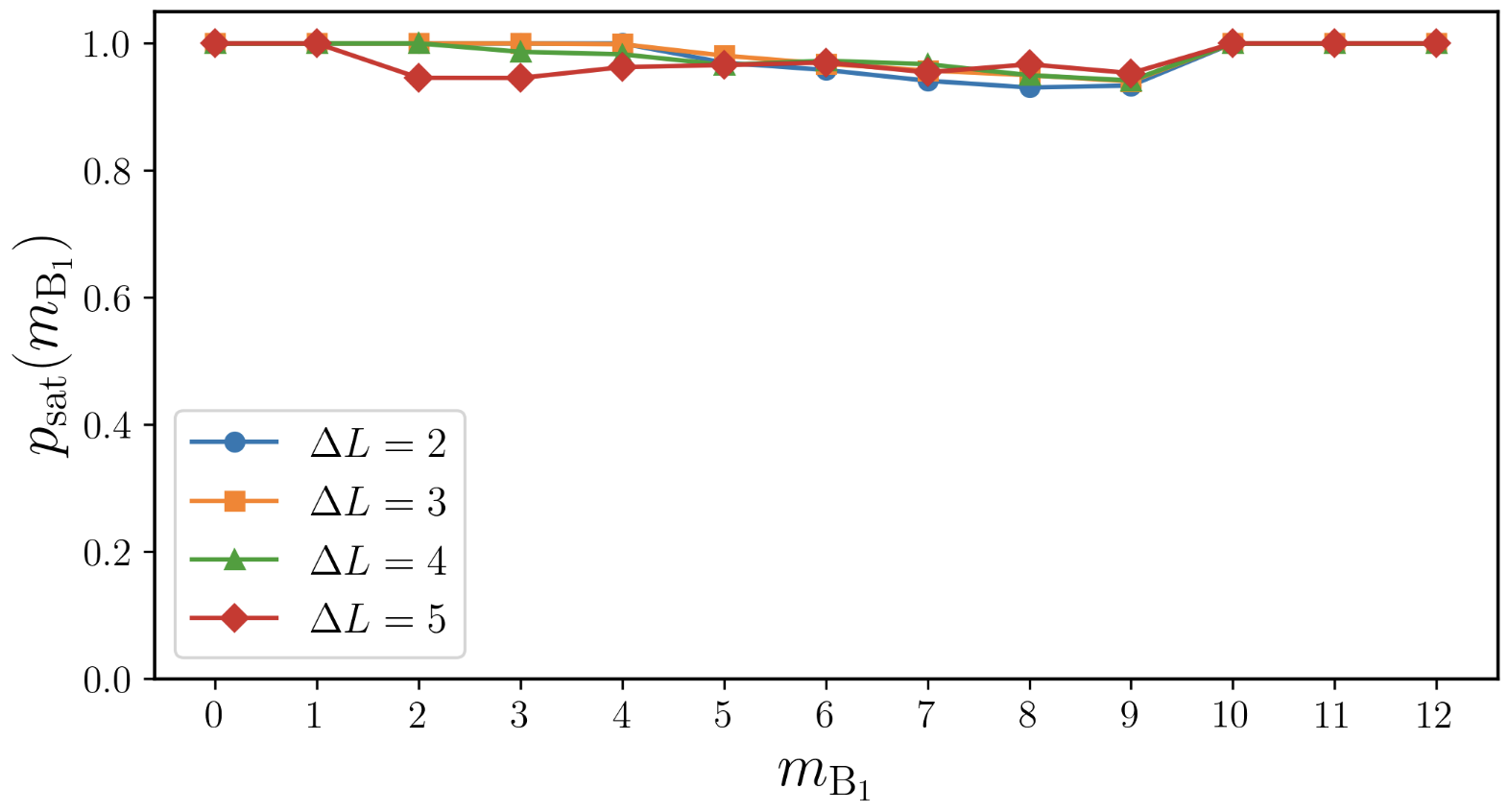}}
\subfloat[$N=14$]{
\includegraphics[width=1.70in]{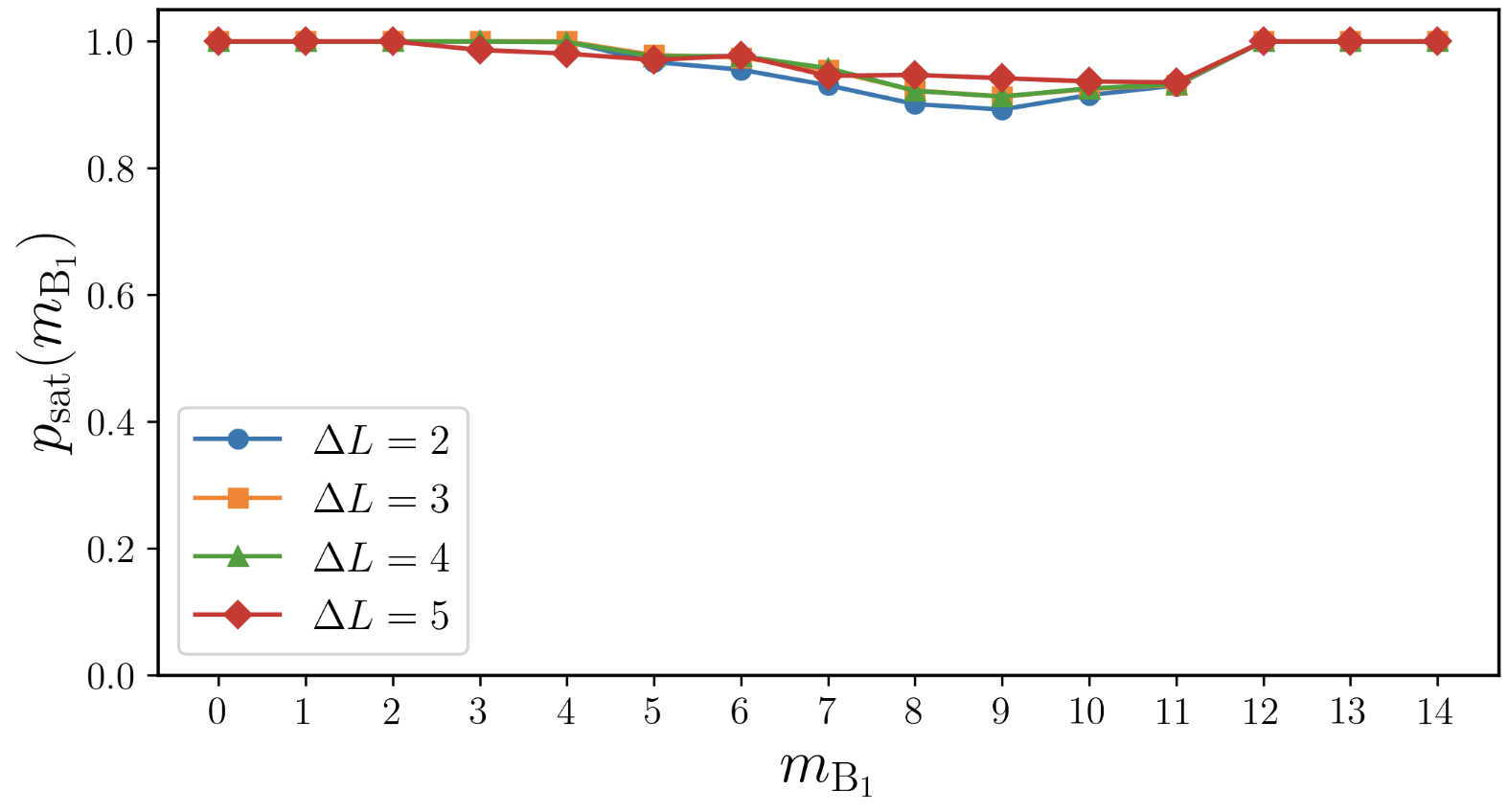}}
\caption{
\textbf{Finite-size scaling of the saturation probability.}
Probability $p_{\mathrm{sat}}$ that the conditional rank saturates the CFT
upper bound of Eq.~(\ref{rankcollapseconjecture}), shown as a function of
the number of orbitals $m_{\mathrm B_1}$ included in the measured subsystem
$\mathrm B_1$ for Laughlin states with
(a) $N=8$, (b) $N=10$, (c) $N=12$, and (d) $N=14$ particles.
The saturation probability is obtained by summing the probabilities of all
measurement outcomes $\mathbf n_{\mathrm B_1}$ for which
$\operatorname{rank}[\rho_{A|\mathbf n_{\mathrm B_1}}]$ equals the
corresponding conditional CFT counting. The point
$m_{\mathrm B_1}=0$ reduces to the conventional Li--Haldane problem and
therefore has $p_{\mathrm{sat}}=0$ or $1$ depending on whether the
finite-size RDM saturates the usual CFT counting. The overall increase of
$p_{\mathrm{sat}}$ with system size is consistent with saturation becoming
generic in the thermodynamic limit.
}
\label{probabilitysaturation}
\end{figure}

\begin{table}[htp]
\centering
\begin{tabular}{c c c c c c}
\hline\hline
$\Delta L_{\mathrm A}$ &
$\mathbf{n}_{\mathrm{B}_1}$ &
$\operatorname{rank}[\rho_{A|\mathbf n_{\mathrm B_1}}]$ &
$\Delta L_{\mathrm{B}_2}$ &
CFT bound &
Saturated? \\
\hline

\multicolumn{6}{c}{$\nu=1/2$ \textbf{Laughlin}}\\
\hline

2 & $\{1\}$  & 1 & 1 & 1 & Yes \\
2 & $\{01\}$ & 2 & 2 & 2 & Yes \\
3 & $\{1\}$  & 2 & 2 & 2 & Yes \\
3 & $\{02\}$ & 1 & 1 & 1 & Yes \\
4 & $\{1\}$  & 3 & 3 & 3 & Yes \\
4 & $\{01\}$ & 4 & 4 & 5 & No \\

\hline

\multicolumn{6}{c}{$\nu=1$ \textbf{Moore--Read}}\\
\hline

2 & $\{1\}$   & 2 & 1 & 2 & Yes \\
2 & $\{11\}$  & 1 & 1 & 1 & Yes \\
3 & $\{11\}$  & 3 & 2 & 3 & Yes \\
3 & $\{021\}$ & 4 & 2 & 4 & Yes \\

\hline\hline
\end{tabular}
\caption{
Representative numerical tests of the conditional CFT rank bound,
Eq.~(\ref{rankcollapseconjecture}), for Laughlin and Moore--Read model
states. In all cases examined, the CRDM rank does not exceed the predicted
CFT counting. Most of the examples shown saturate the bound, while the
$\Delta L_{\mathrm A}=4$, $\mathbf n_{\mathrm B_1}=\{01\}$ Laughlin example
illustrates a finite-size case in which the inequality is strict.
}
\label{tab:rankcollapse}
\end{table}

We first note that the upper bound in Eq.~(\ref{rankcollapseconjecture}) is satisfied for all Laughlin and Moore-Read states, angular-momentum sectors, and measurement outcomes examined numerically. Representative examples are listed in Table~\ref{tab:rankcollapse}. While the bound itself appears to be robust, it need not be saturated at finite system size. We therefore
investigate separately the probability with which a measurement outcome produces a conditional subspace whose rank saturates the CFT bound.

For a fixed system size, angular-momentum sector, and choice of $m_{\mathrm B_1}$, let $\mathcal{N}_{\mathrm{CFT}}^{\mathrm{cond}} (\mathbf n_{\mathrm B_1})$ denote the right-hand side of Eq.~(\ref{rankcollapseconjecture}) for the measurement outcome $\mathbf n_{\mathrm B_1}$. We define the set of saturating outcomes by
\begin{equation}
\mathcal S_{\mathrm{sat}}
= \left\{
\mathbf n_{\mathrm B_1}\,\middle|\,
\operatorname{rank}
\left[
\rho_{A|\mathbf n_{\mathrm B_1}}
\right] = \mathcal{N}_{\text{CFT}}(\mathbf{n}_{\mathrm{B}_{1}}) \right\}
\label{eq:saturating-outcomes}
\end{equation}
The corresponding saturation probability is $
p_{\mathrm{sat}}
= \sum_{\mathbf n_{\mathrm B_1}\in\mathcal S_{\mathrm{sat}}}
p(\mathbf n_{\mathrm B_1})$,
where $
p(\mathbf n_{\mathrm B_1})
= \sum_{\mathbf n_{\mathrm B_2}}
p(\mathbf n_{\mathrm B_1},\mathbf n_{\mathrm B_2})$,
i.e., the total probability weight of all measurement outcomes for which the
conditional rank reaches the CFT upper bound.

Figure~\ref{probabilitysaturation} shows
$p_{\mathrm{sat}}$ as a function of the number of measured orbitals $m_{\mathrm B_1}$ for Laughlin states with $N=8,10,12$, and $14$ particles. As a useful consistency check, $m_{\mathrm B_1}=0$ corresponds to the absence of the additional
$\mathrm B_1|\mathrm B_2$ partition and therefore reduces to the conventional Li--Haldane problem. Consequently, $p_{\mathrm{sat}}(m_{\mathrm B_1}=0)$ is either zero or unity depending on
whether the ordinary Li--Haldane counting is saturated for the chosen finite-size system and $\Delta L_{\mathrm A}$ sector.

The numerical data show a clear tendency toward increasing saturation probability with system size. In particular, for a fixed measurement depth
$m_{\mathrm B_1}$, an increasing fraction of the projected-ensemble weight is carried by outcomes that saturate Eq.~(\ref{rankcollapseconjecture}).
This trend suggests that saturation becomes increasingly generic with increasing particle number and is consistent with $p_{\mathrm{sat}}\rightarrow1$ in the thermodynamic limit.

The finite-size violations themselves also exhibit a characteristic structure. A common non-saturating outcome consists of a contiguous string
of empty orbitals, $\mathbf n_{\mathrm B_1}=\{00\cdots0\}$, adjacent to the orbital cut.
Numerically, the length of the exceptional zero string required to observe non-saturation grows approximately linearly with system size. Thus, although such exceptional outcomes persist at the finite sizes accessible to our numerics, they are displaced progressively farther from the cut as the system grows. This provides an additional indication that finite-size violations of saturation are pushed to increasingly nonlocal measurement patterns in the thermodynamic limit.

Taken together, these observations motivate a stronger asymptotic form of Eq.~(\ref{rankcollapseconjecture}): while the CFT counting provides an upper bound at finite size, the bound appears to become saturated with probability approaching unity as the thermodynamic limit is taken. We emphasize that
our finite-size numerics do not constitute a proof of this statement, but provide strong evidence for this behaviour over all systems accessible to exact calculation.

\subsection{Measurement-induced Entanglement and \\ the Holevo Quantity}

Here, we examine numerically a coarse-grained feature of post-measurement entanglement for Laughlin states captured by the so-called measurement-induced entanglement, and we draw connections to its quantum information theoretic implications.

In the following, we assume that we always work within a fixed $\Delta L_\mathrm{A}$ and $N_\mathrm{A}$ sector of a pure quantum many-body state $|\Psi_{\mathrm{A}\mathrm{B}_1\mathrm{B}_2}\rangle$. Upon a projective measurement on the $\mathrm{B}_1$ region, we obtain a classical-quantum state:
\begin{equation}
\rho_{\mathrm{A}\mathrm{X}_{1}\mathrm{B}_{2}} = \sum_{x} p_{x} \rho_{\mathrm{A}\mathrm{B}_{2}|x}\otimes |x\rangle\langle x|_{\mathrm{X_1}},
\end{equation}
where $\rho_{\mathrm{A}\mathrm{B}_2|x}$ denotes the normalized post-measurement state conditioned on the outcome $x$, and $\mathrm{X_1}$ denotes a classical register holding the outcomes. Since the global state prior to measurement is pure and the measurement on $\mathrm{B}_1$ is projective, each conditional state $\rho_{\mathrm{A}\mathrm{B}_2|x}$ is itself pure and can be written as $\rho_{\mathrm{A}\mathrm{B}_{2}|x} = |\psi_{\mathrm{A}\mathrm{B}_{2}|x}\rangle\langle\psi_{\mathrm{A}\mathrm{B}_{2}|x}|$. One way to capture the quantum correlations conditioned on the classical state of $\mathrm{X}_1$ is the \textit{measurement-induced entanglement} (MIE), defined as~\cite{Lin_probing_2023}
\begin{equation}
    \mathrm{MIE}(\mathrm{A}:\mathrm{B}_2):= \sum_x p_x S_\mathrm{A}(\psi_{\mathrm{A}\mathrm{B}_{2}|x})=\sum_x p_x S(\rho_{\mathrm{A}|x}),
\end{equation}
namely the von Neumann entropy between $\mathrm{A}$ and $\mathrm{B}_2$ on the conditional states $\psi_{\mathrm{A}\mathrm{B}_{2}|x}$, averaged over the measurement outcomes $x$. 
Following the intuition of rank collapse in the projected ensembles, the MIE can be thought of as a function of $m_{\mathrm{B}_1}$, the number of measured orbitals within $\mathrm{B}$, and one expects that the MIE decays as $m_{\mathrm{B}_1}$ increases. The MIE is an entropic quantity that captures the measurement-resolved structure of the entanglement spectrum, and its decay can be understood from a quantum information standpoint with implications on the projected ensembles.

To understand its relevance, we briefly introduce the notion of the \textit{Holevo quantity} in quantum information theory. For a generic state ensemble $\mathcal{E} = \{p_x, \rho_x\}$ of potentially mixed states, the Holevo quantity is defined as
\begin{equation}
    \chi(\mathcal{E}) := S\left(\sum_x p_x \rho_x\right) - \sum_{x}p_x S(\rho_x),
\end{equation}
which quantifies the amount of entropy that is reduced (i.e., information learned) when the classical label $x$ is known. It is established~\cite{Holevo1973} that the Holevo quantity provides an upper bound on the so-called \textit{accessible information} of the ensemble, which we define now.

Consider $\mathcal{M} = \{M_y\}$ a positive operator-valued measurement (POVM) consisting of positive operators $M_y\ge 0$ with the constraint $\sum_y M_y = I$. Then, given a quantum state $\rho$, the probability of measuring $y$ is given by
\begin{equation}
    p(y) = \mathrm{Tr}(\rho M_y),
\end{equation}
while on the ensemble $\mathcal{E} = \{p_x, \rho_x\}$ the conditional probability of measuring $y$ given the state label $x$ is
\begin{equation}
    p(y|x) = \mathrm{Tr}(\rho_x M_y),
\end{equation}
generally different from $p(y)=\sum_x p_x p(y|x)$, the marginal distribution. We may thus define the mutual information between the state indices $x$ and the measurement outcomes $y$
\begin{equation}
    \mathcal{I}(\mathcal{E}:\mathcal{M}):= I(\mathrm{X}:\mathrm{Y}) = S(\mathrm{X}) + S(\mathrm{Y}) - S(\mathrm{XY}),
\end{equation}
where $\mathrm{X}$ and $\mathrm{Y}$ denote classical random variables of state labels in $\mathcal{E}$ and measurement outcomes by $\mathcal{M}$, respectively. This mutual information captures the amount of knowledge learned about $\mathrm{X}$ when the measurement result $\mathrm{Y}$ is obtained, and vice versa.

\begin{figure}[t]
    \centering
    \includegraphics[width=0.6\linewidth]{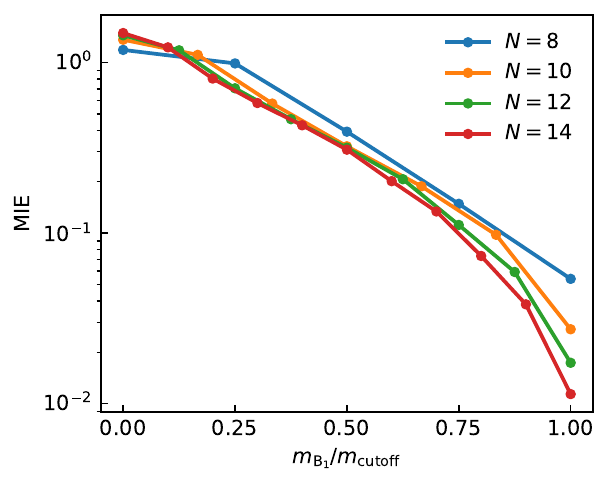}
    \caption{Measurement-induced entanglement (MIE) shown for Laughlin states with $N = 8, 10, 12, 14$ particles overlaid and $\Delta L_\mathrm{A} = 3$ fixed. Here, $m_{\mathrm{B}_1}$ denotes the number of orbitals in $\mathrm{B}_1$, namely the number of measured orbitals within $\mathrm{B}$, and $m_\mathrm{cutoff}$ denotes the largest $m_{\mathrm{B}_1}$ such that the projected states are not completely collapsed to pure states, i.e., the largest $m_{\mathrm{B}_1}$ such that the MIE is non-zero. The plot shows an approximate collapse between different $N$ and also exhibits an approximate exponential decay of MIE with respect to $m_{\mathrm{B}_1}$, confirming our intuition based on rank collapse of the projected ensembles.} 
    \label{fig:mie}
\end{figure}

Then, the accessible information of an ensemble $\mathcal{E}$ is defined as the supremum of that mutual information over all possible measurements
\begin{equation}
    \mathrm{Acc}(\mathcal{E}):= \sup_{\mathcal{M}\in\mathrm{POVM}}\mathcal{I}(\mathcal{E}:\mathcal{M}),
\end{equation}
where $\mathcal{M}$ runs over all possible POVMs, namely the maximum amount of classical information of $\mathrm{X}$ extractable through quantum measurements on the ensemble. This is then upper bounded by the Holevo quantity~\cite{Holevo1973}
\begin{equation}
    \mathrm{Acc}(\mathcal{E}) \le \chi(\mathcal{E}).
\end{equation}

For the projected ensemble of FQH states, we may think of the partial projected ensemble
\begin{equation}
    \mathcal{E}_{m_{\mathrm{B}_1}} = \{ p(\mathbf{n}_{\mathrm{B}_1}), \rho_{\mathrm{A}|\mathbf{n}_{\mathrm{B}_1}} \},
\end{equation}
as progressive unravellings of the RDM $\rho_\mathrm{A}$ with increasing $m_{\mathrm{B}_1}$, where $\rho_\mathrm{A}$ can be understood as $\mathcal{E}_{m_{\mathrm{B}_1} = 0}$. The Holevo information for each of them is
\begin{equation}
\begin{aligned}
    \chi(\mathcal{E}_{m_{\mathrm{B}_1}}) &= S(\rho_\mathrm{A}) - \sum_{\mathbf{n}_{\mathrm{B}_1}} p(\mathbf{n}_{\mathrm{B}_1}) S(\rho_{\mathrm{A}|\mathbf{n}_{\mathrm{B}_1}})
    \\
    &= \mathrm{MIE}_0 - \mathrm{MIE}_{m_{\mathrm{B}_1}},
\end{aligned}
\end{equation}
where we denoted $\mathrm{MIE}_{m_{\mathrm{B}_1}}$ as the MIE of a partition with the size of $\mathrm{B}_1$ being $m_{\mathrm{B}_1}$. Hence, the MIE contains the full quantum information theoretic profile of the Holevo quantity. In Fig.~\ref{fig:mie}, it is shown for Laughlin states that the MIE decays with $\mathrm{m}_{\mathrm{B}_1}$ approximately exponentially.

With this observation, one could understand the measurement of a new orbital in $\mathrm{B}$ from the perspective of the Holevo quantity: an ensemble $\mathcal{E}_{m_{\mathrm{B}_1}}$ is further unravelled into $\mathcal{E}_{m_{\mathrm{B}_1} + 1}$, and the Holevo quantity marginally increases by
\begin{equation}
    \chi(\mathcal{E}_{m_{\mathrm{B}_1} + 1}) - \chi(\mathcal{E}_{m_{\mathrm{B}_1}}) = \mathrm{MIE}_{m_{\mathrm{B}_1}} - \mathrm{MIE}_{m_{\mathrm{B}_1} + 1},
\end{equation}
which by itself decays exponentially. This suggests that the marginal information learned through the measurement of a single orbital decays exponentially with the total number of measured orbitals $m_{\mathrm{B}_1}$.

\subsection{Derivation of Measurement-Resolved Conjecture}

We assume throughout that the total RDM
$\rho_{\mathrm{A}\mathrm{B}_{1}\mathrm{B}_{2}}$ belongs to a fixed
$\Delta L_{\mathrm{A}}$ sector and that the total angular momentum satisfies
\begin{equation}
\Delta L_{\mathrm{tot}}
=
\Delta L_{\mathrm{A}}
+
\Delta L_{\mathrm{B}_{1}}
+
\Delta L_{\mathrm{B}_{2}}
=0.
\end{equation}
For a measurement of subsystem $\mathrm{B}_{1}$ in the orthonormal basis
$\{\ket{x}\}$, the CRDM of subsystem
$\mathrm{B}_{2}$ is
\begin{equation}
\begin{aligned}
\rho_{\mathrm{B}_{2}|x}
=\frac{1}{p(x)}
\mathrm{Tr}_{\mathrm{A}\mathrm{B}_{1}}
[\left(\mathbbm{1}_{\mathrm A}
\otimes|x\rangle\langle x|
\otimes\mathbbm{1}_{\mathrm{B}_{2}}
\right)
\rho_{\mathrm{A}\mathrm{B}_{1}\mathrm{B}_{2}} 
\\ 
\times
\left(\mathbbm{1}_{\mathrm A}
\otimes|x\rangle\langle x|
\otimes\mathbbm{1}_{\mathrm{B}_{2}}
\right)
],
\end{aligned}
\end{equation}
where
\begin{equation}
p(x)
=
\mathrm{Tr}
\left[
\left(
\mathbbm{1}_{\mathrm A}
\otimes
|x\rangle\langle x|
\otimes
\mathbbm{1}_{\mathrm{B}_{2}}
\right)
\rho_{\mathrm{A}\mathrm{B}_{1}\mathrm{B}_{2}}
\right]
\end{equation}
is the probability of obtaining outcome $x$. Averaging over all measurement outcomes gives
\begin{align}
\sum_x p(x)\rho_{\mathrm{B}_{2}|x}
&=
\sum_x
\mathrm{Tr}_{\mathrm{A}\mathrm{B}_{1}}
[
\left(
\mathbbm{1}_{\mathrm A}
\otimes
|x\rangle\langle x|
\otimes
\mathbbm{1}_{\mathrm{B}_{2}}
\right) \\& \times
\rho_{\mathrm{A}\mathrm{B}_{1}\mathrm{B}_{2}}  
\left(
\mathbbm{1}_{\mathrm A}
\otimes
|x\rangle\langle x|
\otimes
\mathbbm{1}_{\mathrm{B}_{2}}
\right)
]
\nonumber\\
&=
\mathrm{Tr}_{\mathrm{A}\mathrm{B}_{1}}
\left[
\rho_{\mathrm{A}\mathrm{B}_{1}\mathrm{B}_{2}}
\right]
\nonumber\\
&=
\rho_{\mathrm{B}_{2}},
\label{eq:average-conditional-rdm}
\end{align}
where we have used the completeness relation $\sum_x |x\rangle\langle x|=\mathbbm{1}_{\mathrm{B}_{1}}$. Since every term in Eq.~\eqref{eq:average-conditional-rdm} is positive
semidefinite, it follows that
$
\operatorname{supp}
\left(
\rho_{\mathrm{B}_{2}|x}
\right)
\subseteq
\operatorname{supp}
\left(
\rho_{\mathrm{B}_{2}}
\right)$,
and hence
\begin{equation}
\operatorname{rank}
\left(
\rho_{\mathrm{B}_{2}|x}
\right)
\leq
\operatorname{rank}
\left(
\rho_{\mathrm{B}_{2}}
\right)
\end{equation}
for every outcome $x$ with $p(x)>0$.

We now restrict the sum to measurement outcomes that produce conditional
states in a fixed angular-momentum sector $\Delta L_{\mathrm{B}_{2}}=\ell$.
Because both $\Delta L_{\mathrm A}$ and $\Delta L_{\mathrm{tot}}=0$ are
fixed, conservation of angular momentum implies that the corresponding
measurement outcomes on subsystem $\mathrm{B}_{1}$ satisfy $\Delta L_{\mathrm{B}_{1}}(x)
=
-\Delta L_{\mathrm A}-\ell$. Let
\begin{equation}
X_{\ell}
=
\left\{
x:
\Delta L_{\mathrm{B}_{1}}(x)
=
-\Delta L_{\mathrm A}-\ell
\right\}
\end{equation}
denote the set of all measurement outcomes belonging to this
$\mathrm{B}_{1}$ angular-momentum sector. We assume that the states
$\{\ket{x}\}_{x\in X_{\ell}}$ form a complete orthonormal basis of that
sector. Consequently,
\begin{equation}
\sum_{x\in X_{\ell}}
|x\rangle\langle x|
=
\Pi_{\mathrm{B}_{1}}^{-\Delta L_{\mathrm A}-\ell},
\end{equation}
where $\Pi_{\mathrm{B}_{1}}^{m}$ denotes the projector onto the
$\Delta L_{\mathrm{B}_{1}}=m$ sector.
Summing the unnormalised conditional density matrices over these outcomes
gives
\begin{equation}
\begin{aligned}
\sum_{x\in X_{\ell}}
p(x)\rho_{\mathrm{B}_{2}|x}
&=
\sum_{x\in X_{\ell}}
\mathrm{Tr}_{\mathrm{A}\mathrm{B}_{1}}
[
\left(
\mathbbm{1}_{\mathrm A}
\otimes
|x\rangle\langle x|
\otimes
\mathbbm{1}_{\mathrm{B}_{2}}
\right) \\ &\times
\rho_{\mathrm{A}\mathrm{B}_{1}\mathrm{B}_{2}}
\left(
\mathbbm{1}_{\mathrm A}
\otimes
|x\rangle\langle x|
\otimes
\mathbbm{1}_{\mathrm{B}_{2}}
\right)
]
\nonumber\\
&=
\mathrm{Tr}_{\mathrm{A}\mathrm{B}_{1}}
\left[
\left(
\mathbbm{1}_{\mathrm A}
\otimes
\Pi_{\mathrm{B}_{1}}^{-\Delta L_{\mathrm A}-\ell}
\otimes
\mathbbm{1}_{\mathrm{B}_{2}}
\right)
\rho_{\mathrm{A}\mathrm{B}_{1}\mathrm{B}_{2}}
\right].
\label{eq:restricted-outcome-sum}
\end{aligned}
\end{equation}

On the support of
$\rho_{\mathrm{A}\mathrm{B}_{1}\mathrm{B}_{2}}$, the constraints
$\Delta L_{\mathrm{tot}}=0
\qquad$ and $\qquad
\Delta L_{\mathrm A}
$ is fixed imply $\Delta L_{\mathrm{B}_{1}}
=
-\Delta L_{\mathrm A}-\Delta L_{\mathrm{B}_{2}}$.
It follows that, on this support, projecting subsystem $\mathrm{B}_{1}$
onto the sector
$\Delta L_{\mathrm{B}_{1}}=-\Delta L_{\mathrm A}-\ell$
is equivalent to projecting subsystem $\mathrm{B}_{2}$ onto the sector
$\Delta L_{\mathrm{B}_{2}}=\ell$. Therefore,
\begin{align}
\sum_{x\in X_{\ell}}
p(x)\rho_{\mathrm{B}_{2}|x}
&=
\Pi_{\mathrm{B}_{2}}^{\ell}
\,
\mathrm{Tr}_{\mathrm{A}\mathrm{B}_{1}}
\left[
\rho_{\mathrm{A}\mathrm{B}_{1}\mathrm{B}_{2}}
\right]
\,
\Pi_{\mathrm{B}_{2}}^{\ell}
\nonumber\\
&=
\Pi_{\mathrm{B}_{2}}^{\ell}
\rho_{\mathrm{B}_{2}}
\Pi_{\mathrm{B}_{2}}^{\ell}
\nonumber\\
&\equiv
\rho_{\mathrm{B}_{2}}(\ell).
\label{eq:restricted-average-sector}
\end{align}
Thus, the average over all measurement outcomes that produce a conditional
state with fixed
$\Delta L_{\mathrm{B}_{2}}=\ell$
is exactly the corresponding angular-momentum block of the ordinary reduced
density matrix:
\begin{equation}
\sum_{x\in X_{\ell}}
p(x)\rho_{\mathrm{B}_{2}|x}
=
\rho_{\mathrm{B}_{2}}(\ell).
\end{equation}

Since all operators in this sum are positive semidefinite, for every
$x\in X_{\ell}$ with $p(x)>0$ we have
$\operatorname{supp}
\left(
\rho_{\mathrm{B}_{2}|x}
\right)
\subseteq
\operatorname{supp}
\left(
\rho_{\mathrm{B}_{2}}(\ell)
\right)$.
Consequently,
\begin{equation}
\operatorname{rank}
\left(
\rho_{\mathrm{B}_{2}|x}
\right)
\leq
\operatorname{rank}
\left(
\rho_{\mathrm{B}_{2}}(\ell)
\right).
\end{equation}
Therefore, any upper bound on the rank of
$\rho_{\mathrm{B}_{2}}(\ell)$ supplied by the ordinary Li--Haldane
conjecture also provides an upper bound on the rank of every conditional
RDM
$\rho_{\mathrm{B}_{2}|x}$
in the same $\Delta L_{\mathrm{B}_{2}}=\ell$ sector. In this sense, the
generalised Li--Haldane rank bound follows from the ordinary Li--Haldane
rank bound, provided that the sum is taken over a complete measurement
basis of the corresponding $\mathrm{B}_{1}$ angular-momentum sector.

\subsection{Rank rigidity}
\label{sec:rank-rigidity}

\subsubsection{Numerical observations}

\begin{figure}[htp]
\centering
\includegraphics[width=3.4in]
{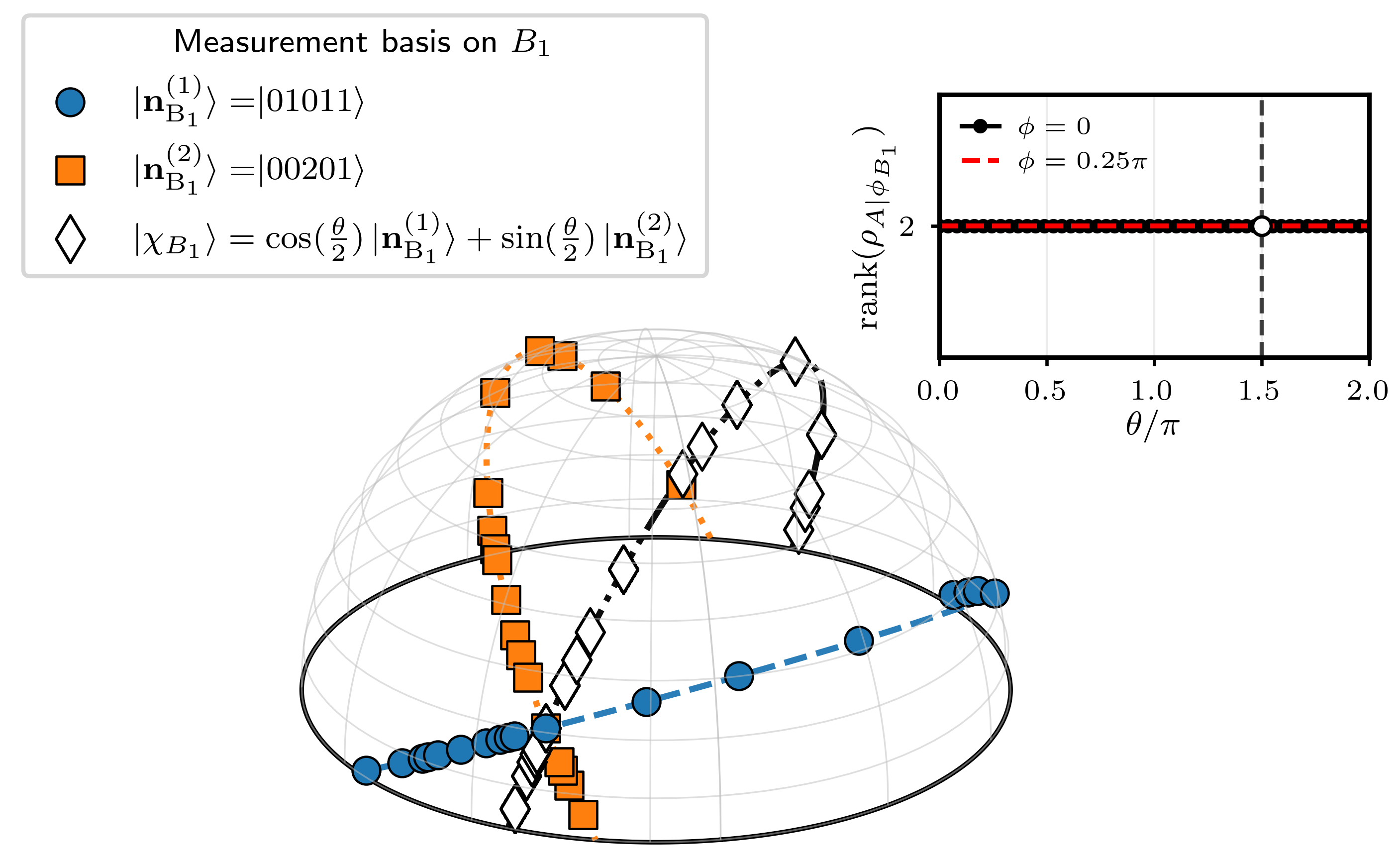}
\caption{
Interpolation between two occupation-number measurement outcomes
$\ket{\mathbf n_{\mathrm B_1}^{(1)}}$ and
$\ket{\mathbf n_{\mathrm B_1}^{(2)}}$ in the
$\Delta L_{\mathrm A}=3$ sector of the bosonic Laughlin state, obtained by
conditioning on
$\ket{\chi_{\mathrm B_1}(\theta,\phi)}$ defined in
Eq.~(\ref{eq:rank-rigidity-interpolation}).
As the measurement state is continuously deformed, the corresponding
rank-collapsed subspace evolves nontrivially between the two endpoint
configurations while preserving its rank. The inset shows the rank of the
CRDM $\rho_{A|\chi_{\mathrm B_1}(\theta,\phi)}$ as a function of $\theta$
for representative values of $\phi$. For the interpolation shown in the main
panel, a special value $\theta=\theta_{\mathrm c}$ produces a collapse point
at which all trajectories except the fixed point coalesce, while the two
distinct points continue to span a rank-$2$ subspace.
}
\label{RankRigidityDeltaL=3}
\end{figure}

We characterize the basis dependence of the rank-collapsed subspaces by considering coherent interpolations between two occupation-number configurations of $\mathrm B_1$,
\begin{equation}
\ket{\chi_{\mathrm B_1}(\theta,\phi)}
=
\cos\left(\frac{\theta}{2}\right)
\ket{\mathbf n_{\mathrm B_1}^{(1)}}
+
\sin\left(\frac{\theta}{2}\right)
\ee^{\ii\phi}
\ket{\mathbf n_{\mathrm B_1}^{(2)}} ,
\label{eq:rank-rigidity-interpolation}
\end{equation}
and study the subspace of the PE conditioned on the
measurement outcome $\chi_{\mathrm B_1}(\theta,\phi)$. Equivalently, we
consider the rank of the corresponding conditional reduced density matrix
$\rho_{A|\chi_{\mathrm B_1}(\theta,\phi)}$.

Figure~\ref{RankRigidityDeltaL=3} shows a representative example in the
$\Delta L_{\mathrm A}=3$ sector of the bosonic Laughlin state. Although the
geometry of the conditioned subspace evolves nontrivially as the measurement
state is continuously rotated between the two endpoint configurations, its
rank remains unchanged throughout the interpolation. We refer to this
basis-independent persistence of the conditional rank as
\textit{rank rigidity}.

For interpolations between rank-$2$ conditional subspaces, the evolution
exhibits additional geometric structure. In particular, we frequently find a
special interpolation angle $\theta=\theta_{\mathrm c}$ at which all
trajectories, with the exception of a single fixed point (lying at the intersection of the two rank-2 subspaces), coalesce at a
second point. We refer to this second point as the \textit{collapse point}.
Importantly, this geometric collapse does not necessarily imply a reduction of the CRDM
rank: the fixed point and collapse point may remain distinct and continue to span
a rank-$2$ subspace.

Numerically, we observe three qualitatively distinct classes of rank-$2$
interpolations:
\begin{enumerate}
    \item \textit{Non-overlapping subspaces with distinct fixed and collapse
    points.} The two  rank-$2$ subspaces do not coincide, while the
    interpolation contains both a fixed point and a distinct collapse point.

    \item \textit{Overlapping subspaces with coincident fixed and collapse
    points.} The rank-$2$ subspaces overlap, and the fixed point
    coincides with the collapse point.

    \item \textit{Non-overlapping subspaces without fixed or collapse
    points.} The endpoint subspaces are distinct, but the interpolation
    exhibits neither a common fixed point nor a collective collapse point.
\end{enumerate}
These observations show that rank rigidity constrains the conditioned
subspaces more strongly than their dimension alone: under coherent rotations
of the measurement basis, the subspaces undergo a structured deformation
within the PE while preserving their rank. In the following
section, we focus on the first class, i.e. non-overlapping rank-$2$ subspaces with
a distinct fixed point and collapse point, for which we derive an exact
analytic mapping between the two rank-collapsed manifolds.

\subsubsection{Exact Mapping Between Rank-2 Conditional Subspaces}
We define the following measurement bases, $\ket{\alpha}\equiv\ket{\mathbf{n}_{\mathrm{B}_1}^{(1)}}$, $\ket{\beta}\equiv\ket{\mathbf{n}_{\mathrm{B}_1}^{(2)}}$, $\ket{\gamma}\equiv\ket{\chi_{\mathrm{B}_1}(\theta)} = 
\cos(\frac{\theta}{2})\ket{\mathbf{n}_{\mathrm{B}_1}^{(1)}} +\sin(\frac{\theta}{2})\ket{\mathbf{n}_{\mathrm{B}_1}^{(2)}}$. Measuring with $\ket{\gamma}$, we obtain a set of states $\ket{\psi_{\gamma}(\mathbf{n}_{\mathrm{B}_{2}})}$ which belong to the full rank-3 space (e.g. $\Delta L_{\mathrm{A}}=3$ sector of Laughlin or $\Delta L_{\mathrm{A}}=2$ sector of Moore-Read). These states can then be written in terms of post measurement states of $\alpha$ and $\beta$ as
\begin{equation}
\label{rankconservationeq}
\begin{aligned}
\sqrt{p_{\gamma}(\mathbf{n}_{\mathrm{B}_{2}})} \ket{\psi_{\gamma}(\mathbf{n}_{\mathrm{B}_{2}})} = \sqrt{p_{\alpha}(\mathbf{n}_{\mathrm{B}_{2}})} \cos(\theta/2) \ket{\psi_{\alpha}(\mathbf{n}_{\mathrm{B}_{2}})} \\ + \sqrt{p_{\beta}(\mathbf{n}_{\mathrm{B}_{2}})} \sin(\theta/2) \ket{\psi_{\beta}(\mathbf{n}_{\mathrm{B}_{2}})},
\end{aligned}
\end{equation}
where $p_{\gamma}(\mathbf{n}_{\mathrm{B}_{2}}) = p_{\alpha}(\mathbf{n}_{\mathrm{B}_{2}})\cos^{2}(\theta/2) + p_{\beta}(\mathbf{n}_{\mathrm{B}_{2}})\sin^{2}(\theta/2) + \sqrt{p_{\alpha}(\mathbf{n}_{\mathrm{B}_{2}}) p_{\beta}(\mathbf{n}_{\mathrm{B}_{2}})}\sin(\theta)\braket{\psi_{\alpha}(\mathbf{n}_{\mathrm{B}_{2}}) |\psi_{\beta}(\mathbf{n}_{\mathrm{B}_{2}})}$. The vectors coming from measurements of $\alpha$ and $\beta$ are assumed to span two different rank-2 subspaces of the full rank-3 space. We note that the simplest solution of Eq.~(\ref{rankconservationeq}) occurs when both sets of vectors span the same rank-2 subspace. Indeed, this situation is occasionally observed in the numerics. But now we will see what the property of the transformation between subspaces is if we maintain that the two rank-2 subspaces are different, and we assume the rank rigidity property telling us that $\gamma$ vectors are also spanning a rank-2 subspace. We will also assume that there exists a common vector in both the $\alpha$ and $\beta$ subspaces, which we denote by $\ket{e_{0}}$. If this vector is itself realised as a projected-ensemble state in both subspaces, it follows directly from Eq.~(\ref{rankconservationeq}) that it is a fixed point of the interpolation. This case is also frequently seen in numerics. 

Expanding both $\alpha$ and $\beta$ subspaces in the basis containing the fixed point, we can write $\ket{\psi_{\alpha}(\mathbf{n}_{\mathrm{B}_{2}})} = a_{0}(\mathbf{n}_{\mathrm{B}_{2}})\ket{e_{0}} + a (\mathbf{n}_{\mathrm{B}_{2}})\ket{e_{\alpha}}$,$\ket{\psi_{\beta}(\mathbf{n}_{\mathrm{B}_{2}})} = b_{0}(\mathbf{n}_{\mathrm{B}_{2}})\ket{e_{0}} + b (\mathbf{n}_{\mathrm{B}_{2}})\ket{e_{\beta}}$. Defining $\lambda(\theta, \mathbf{n}_{\mathrm{B}_{2}}) = \sqrt{\frac{p_{\beta}(\mathbf{n}_{\mathrm{B}_{2}})}{p_{\alpha}(\mathbf{n}_{\mathrm{B}_{2}})}}\tan(\theta/2)$, we get
\begin{equation}
\label{gammaexpressionrankrigidity}
\begin{aligned}
\ket{\psi_{\gamma}(\mathbf{n}_{\mathrm{B}_{2}})} \propto (a_{0}(\mathbf{n}_{\mathrm{B}_{2}}) + \lambda(\theta, \mathbf{n}_{\mathrm{B}_{2}}) b_{0}(\mathbf{n}_{\mathrm{B}_{2}}))\ket{e_{0}} \\+ (a(\mathbf{n}_{\mathrm{B}_{2}}) + \alpha\lambda(\theta, \mathbf{n}_{\mathrm{B}_{2}}) b(\mathbf{n}_{\mathrm{B}_{2}}))\ket{e_{\alpha}} \\ + \sqrt{1-\alpha^{2}}\lambda(\theta, \mathbf{n}_{\mathrm{B}_{2}}) b(\mathbf{n}_{\mathrm{B}_{2}}))\ket{n_{\alpha}}, 
\end{aligned}
\end{equation}
where we have decomposed $\ket{e_{\beta}} = \alpha \ket{e_{\alpha}}+\sqrt{1-\alpha^{2}}\ket{n_{\alpha}}$, $\alpha = \braket{e_{\alpha}|e_{\beta}}$, and $\ket{n_{\alpha}}$ is the vector orthogonal to the $\alpha$ subspace. Since $\ket{e_{0}}$ is always present in the $\gamma$ subspace due to fixed-point solution, rank rigidity requires that the component of $\ket{\psi_{\gamma}(\mathbf{n}_{\mathrm{B}_{2}})}$ orthogonal to $\ket{e_{0}}$ points in the same direction independent of $\mathbf{n}_{\mathrm{B}_{2}}$. Equivalently, the ratio of the coefficients of $\ket{e_{\alpha}}$ and $\ket{n_{\alpha}}$ must be independent of $\mathbf{n}_{\mathrm{B}_{2}}$. Therefore, 
\begin{equation}
\frac{\lambda(\theta, \mathbf{n}_{\mathrm{B}_{2}})b(\mathbf{n}_{\mathrm{B}_{2}})\sqrt{1-\alpha^{2}}}{a(\mathbf{n}_{\mathrm{B}_{2}}) + \alpha \lambda(\theta,\mathbf{n}_{\mathrm{B}_{2}})b(\mathbf{n}_{\mathrm{B}_{2}})} = f(\theta,\alpha),
\end{equation}
for some function $f(\theta, \alpha)$. This directly implies that
\begin{equation}
\lambda(\theta, \mathbf{n}_{\mathrm{B}_{2}}) = g(\theta) \frac{a(\mathbf{n}_{\mathrm{B}_{2}})}{b(\mathbf{n}_{\mathrm{B}_{2}})}.
\end{equation}
From our definition of $\lambda$, we get $g(\theta) = \frac{1}{C} \tan(\theta/2)$, for $C$ some constant, and the important constraint relating the coefficients of the $\alpha$ and $\beta$ vectors to their associated probabilities
\begin{equation}
\sqrt{\frac{p_{\alpha}(\mathbf{n}_{\mathrm{B}_{2}})}{p_{\beta}(\mathbf{n}_{\mathrm{B}_{2}})}}\frac{a(\mathbf{n}_{\mathrm{B}_{2}})}{b(\mathbf{n}_{\mathrm{B}_{2}})} = C. 
\end{equation}
This relation constitutes a strong constraint linking the probabilities and geometric structure of the two rank-2 subspaces. This then also gives us an expression for $f(\theta,\alpha)$:
\begin{equation}
f(\theta,\alpha) = \frac{\tan(\theta/2) \sqrt{1-\alpha^{2}}}{C+\alpha \tan(\theta/2)}.
\end{equation}
After plugging these constraints into Eq.~(\ref{gammaexpressionrankrigidity}), we find the final expression for the interpolation between the two rank-2 subspaces with a fixed point: 
\begin{equation}
\begin{aligned}
\ket{\psi_{\gamma}(\mathbf{n}_{\mathrm{B}_{2}})}\propto \left(C\frac{a_{0}(\mathbf{n}_{\mathrm{B}_{2}})}{a(\mathbf{n}_{\mathrm{B}_{2}})}+\tan(\theta/2)\frac{b_{0}(\mathbf{n}_{\mathrm{B}_{2}})}{b(\mathbf{n}_{\mathrm{B}_{2}})}\right) \ket{e_{0}} \\ + \left(C+\alpha \tan(\theta/2)\right)\ket{e_{\alpha}} + \tan(\theta/2)\sqrt{1-\alpha^{2}}\ket{n_{\alpha}}.
\end{aligned}
\end{equation}

This is as far as the rank rigidity assumption alone takes us analytically. Numerically, we observe an additional phenomenon. Namely, in all examples exhibiting both rank rigidity and a fixed point, there exists a special value $\theta=\theta^{\star}$ for which all projected-ensemble states, except the fixed point itself, collapse onto a single point. We refer to this as the \textit{collapse point}. At present, we do not have a first-principles derivation of this condition and therefore regard it as an empirical observation motivated by the numerics. Once assumed to exist, the collapse point condition in our analytics is $C\frac{a_{0}(\mathbf{n}_{\mathrm{B}_{2}})}{a(\mathbf{n}_{\mathrm{B}_{2}})}+\tan(\theta/2)\frac{b_{0}(\mathbf{n}_{\mathrm{B}_{2}})}{b(\mathbf{n}_{\mathrm{B}_{2}})} = D$, for some constant $D$. Writing $a(\mathbf{n}_{\mathrm{B}_{2}}) = \sin(A(\mathbf{n}_{\mathrm{B}_{2}})/2)$ and $b(\mathbf{n}_{\mathrm{B}_{2}}) = \sin(B(\mathbf{n}_{\mathrm{B}_{2}})/2)$, the collapse point assumption gives us an exact mapping between points in the two rank-2 subspaces with rank rigidity:
\begin{equation}
\label{rankmapping}
\cot(B(\mathbf{n}_{\mathrm{B}_{2}})/2) = \frac{D-C\cot(A(\mathbf{n}_{\mathrm{B}_{2}})/2)}{\tan(\theta^{\star}/2)}.
\end{equation}
We can check the consistency of this transformation by considering its inverse. Starting the interpolation at $\beta$, we have 
\begin{equation}
\cot(A(\mathbf{n}_{\mathrm{B}_{2}})/2) = \frac{E-(1/C)\cot(B(\mathbf{n}_{\mathrm{B}_{2}})/2)}{\tan(\frac{\pi -\theta^{\star}}{2})},
\end{equation}
for some new constant $E$. 
Rearranging Eq.~(\ref{rankmapping}), we get
\begin{equation}
\cot(A(\mathbf{n}_{\mathrm{B}_{2}})/2) = (D/C)-(1/C) \tan(\theta^{\star}/2)\cot(B(\mathbf{n}_{\mathrm{B}_{2}})/2).
\end{equation}
These two transformations are consistent, as $\tan(\theta^{\star}/2) = \cot(\frac{\pi-\theta^{\star}}{2})$, and where we identify $E\tan(\theta^{\star}/2) = D/C$. We also note that Eq.~(\ref{rankmapping}) belongs to a class of real projective (Möbius) mappings.

Thus, under the assumptions of rank rigidity, a fixed point, and the empirically observed collapse-point condition, we obtain an exact analytic mapping between points belonging to distinct rank-2 subspaces.  Understanding the microscopic origin of the collapse-point condition, rather than imposing it as an additional empirical input, remains an interesting direction for future work. 

\subsection{Coulomb ground state: edge projection and UV broadening}
\label{sec:coulomb-supp}

\begin{figure}[htp]
\subfloat[]{
\includegraphics[width=1.75in]{spectrum_coulomb_with_model_inset.pdf}}
\subfloat[]{
\includegraphics[width=1.75in]{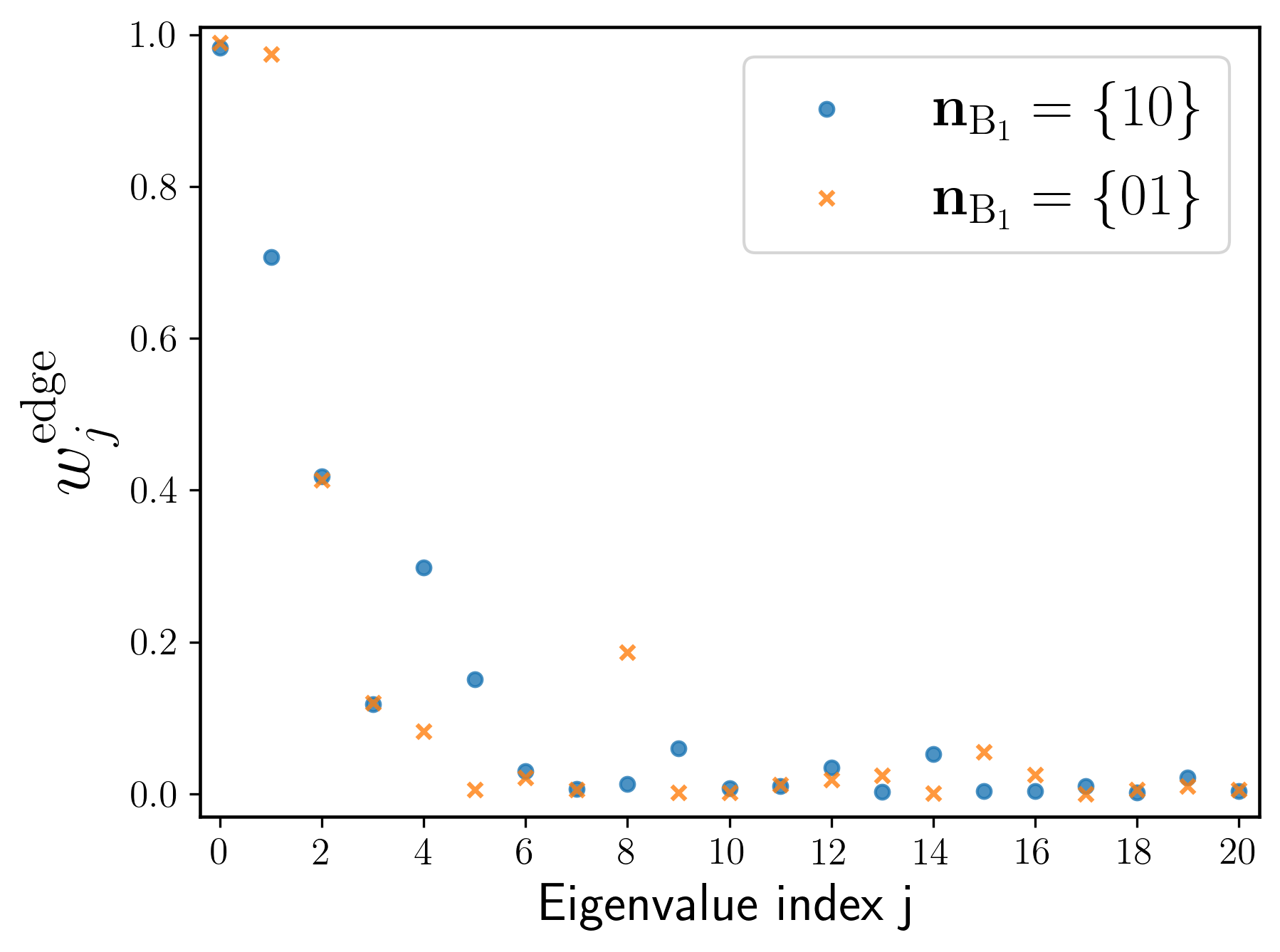}}
\\
\subfloat[]{
\includegraphics[width=1.68in]{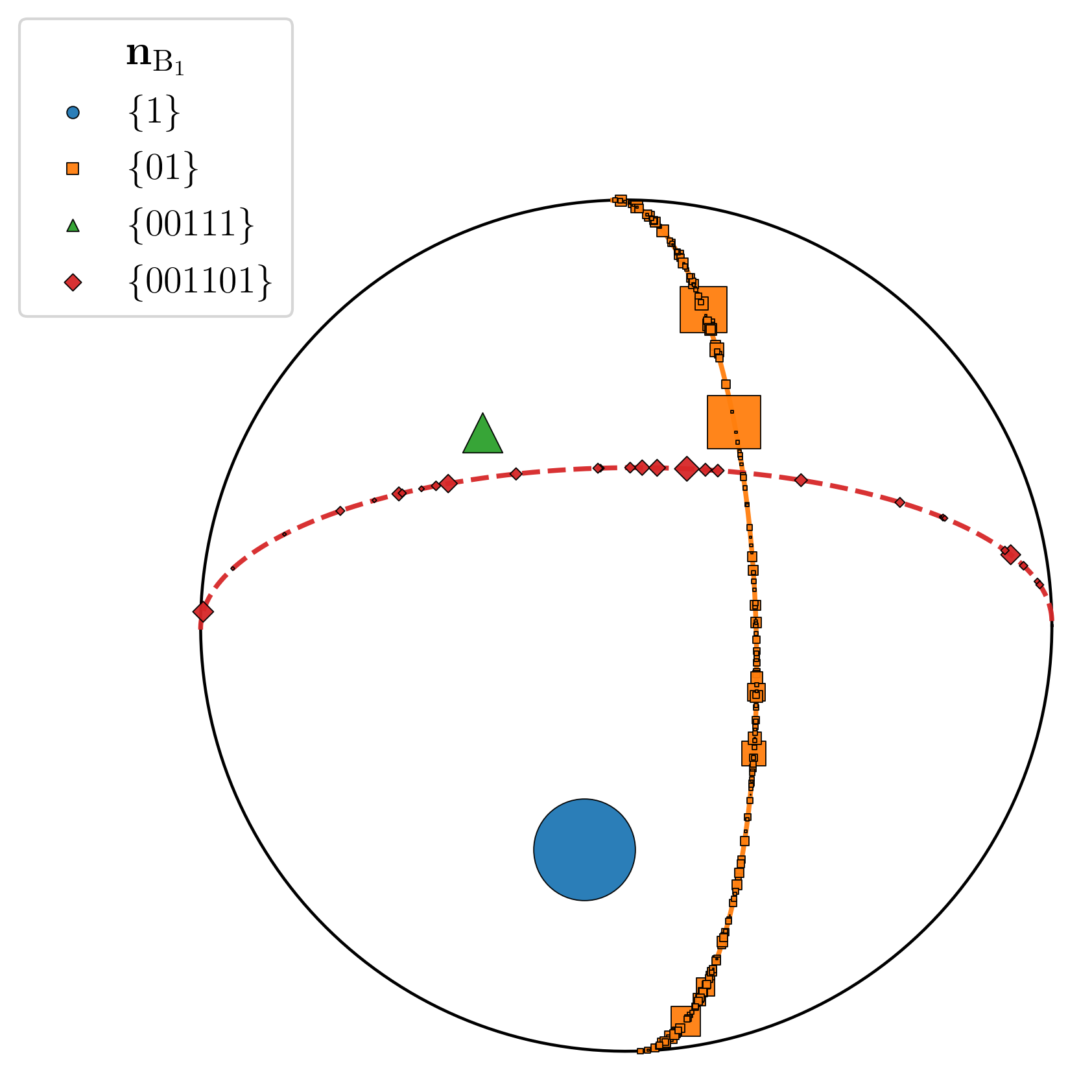}}
\subfloat[]{
\includegraphics[width=1.75in]{DeltaL=2_PE_N_16_q_1_coulomb_CRDM_states_disk.png}}
\caption{
(a) Conditional entanglement spectrum for
$\mathbf n_{\mathrm B_1}=\{1\}$. The universal low-lying sector exhibits the
CFT counting predicted by Eq.~(\ref{rankcollapseconjecture}) and is separated
from the non-universal UV sector by an entanglement gap. The inset shows the
corresponding spectrum of the fermionic Moore-Read model state.
(b) Edge weight $w_j^{\mathrm{edge}}$ of the CRDM eigenstates within the
universal edge subspace of the full Coulomb RDM. The eigenstates spanning
$\mathcal{H}_{A|\mathbf n_{\mathrm B_1}}^{\mathrm{CRDM}}$ have near-unit
edge weight, whereas the UV eigenstates retain a smaller but finite overlap.
(c) PE of the fermionic Moore--Read model state in the $\Delta L_{\mathrm A}=2$ sector. The post-measurement states lie on the conditional rank-collapse manifolds
$\mathcal{H}_{A|\mathbf n_{\mathrm B_1}}^{\mathrm{CRDM}}$. (d) Coulomb PE projected onto the universal edge manifold of the full RDM. The low-lying conditional subspaces $\mathcal{H}_{A|\mathbf n_{\mathrm B_1}}^{\mathrm{CRDM}}$ reproduce not only
the dimensions but also the relative geometric organization of their
Moore--Read counterparts in panel (c). The individual post-measurement states
cluster around these subspaces with a small broadening arising from the finite
edge weight of the UV CRDM sector shown in panel (b).
}
\label{CoulombFigureNew}
\end{figure}

In the main text, we showed that the low-lying sector of the Coulomb CRDMs
obeys the same conditional CFT counting as the Moore-Read model state and that the corresponding conditional subspaces retain essentially the same organization within the PE. Here we examine this correspondence in more detail and clarify the origin of the small broadening away from the ideal rank-collapse manifolds.

Figure~\ref{CoulombFigureNew}(a) shows a representative conditional
entanglement spectrum for
$\mathbf n_{\mathrm B_1}=\{1\}$. As discussed in the main text, the
low-lying universal levels are separated from the non-universal UV sector by an entanglement gap and span the conditional edge subspace
$\mathcal{H}_{A|\mathbf n_{\mathrm B_1}}^{\mathrm{CRDM}}$. To determine how
this subspace is embedded within the conventional Li--Haldane edge manifold,
we define the projector onto the universal sector of the full Coulomb RDM,
\begin{equation}
P_{\mathrm{edge}}^{\mathrm{RDM}}
=
\sum_{j=1}^{\mathcal N_{\mathrm{CFT}}}
\ket{\lambda_j^{\mathrm{RDM}}}
\bra{\lambda_j^{\mathrm{RDM}}},
\label{eq:coulomb-edge-projector}
\end{equation}
where the sum runs over the $\mathcal N_{\mathrm{CFT}}$ RDM eigenstates below
the Li--Haldane entanglement gap. For each CRDM eigenstate we then define its
weight within this edge manifold as
\begin{equation}
w_j^{\mathrm{edge}}
=
\bra{\lambda_j^{\mathrm{CRDM}}}
P_{\mathrm{edge}}^{\mathrm{RDM}}
\ket{\lambda_j^{\mathrm{CRDM}}}.
\label{eq:crdm-edge-weight}
\end{equation}

As shown in Fig.~\ref{CoulombFigureNew}(b), the low-lying CRDM eigenstates
have $w_j^{\mathrm{edge}}$ close to unity, corresponding to the CRDM counting expected from (\ref{rankcollapseconjecture}). The conditional subspaces
$\mathcal{H}_{A|\mathbf n_{\mathrm B_1}}^{\mathrm{CRDM}}$ therefore lie, to
very good approximation, within the edge manifold of the full Coulomb RDM.
By contrast, eigenstates above the conditional entanglement gap have a
smaller edge weight. Importantly, however, the UV CRDM sector retains a finite overlap with the RDM edge manifold.
The universal and UV conditional sectors are therefore strongly, but not
perfectly, separated by the RDM edge projection.

This finite overlap provides a simple explanation for the broadening of the
Coulomb PE. A generic post-measurement state contains components along both the low-lying and UV eigenstates of its corresponding
CRDM. For visualization, we project each post-measurement state onto the
universal RDM edge manifold and normalize,
\begin{equation}
\ket{\widetilde{\psi}_{A}(\mathbf n_{\mathrm B})}
=
\frac{
P_{\mathrm{edge}}^{\mathrm{RDM}}
\ket{\psi_A(\mathbf n_{\mathrm B})}
}{
\sqrt{
\bra{\psi_A(\mathbf n_{\mathrm B})}
P_{\mathrm{edge}}^{\mathrm{RDM}}
\ket{\psi_A(\mathbf n_{\mathrm B})}
}
}.
\label{eq:projected-coulomb-pe}
\end{equation}
Because the low-lying CRDM eigenstates have near-unit edge weight, their
geometry is essentially unchanged by this projection. The finite edge weight
of the UV CRDM eigenstates, however, means that a small non-universal
component also survives, displacing individual projected states away from
the ideal conditional subspaces.

This is illustrated directly in Figs.~\ref{CoulombFigureNew}(c) and
\ref{CoulombFigureNew}(d). For the fermionic Moore-Read model state,
Fig.~\ref{CoulombFigureNew}(c), the projected states lie exactly on the
rank-collapse manifolds associated with the conditional subspaces
$\mathcal{H}_{A|\mathbf n_{\mathrm B_1}}^{\mathrm{CRDM}}$. Remarkably, for
the Coulomb ground state, Fig.~\ref{CoulombFigureNew}(d), the corresponding
low-lying CRDM subspaces exhibit essentially the same relative geometric
organization within the edge manifold. That is, not only their dimensions,
but also their relative positions with respect to one another closely match
those of the Moore--Read model state. The individual post-measurement Coulomb
states cluster around these subspaces but acquire a finite transverse spread,
consistent with the residual UV edge weight observed in Fig.~\ref{CoulombFigureNew}(b).

Thus, the deviation of the Coulomb PE from exact rank collapse does not arise
from a significant rearrangement of the universal conditional subspaces.
Rather, the conditional subspaces retain, to high accuracy, both the dimensions and the relative geometric organization of their Moore-Read counterparts. The observed broadening instead results from the small UV
component that survives projection onto the RDM edge manifold.

\end{document}